\documentclass[11pt]{article}
\usepackage{epsfig,times}
\usepackage{graphicx}
\usepackage{natbib}
\usepackage{amsmath}
\bibpunct{(}{)}{,}{a}{,}{,}

\def\tr{{\mathrm T}}  
\newcommand{\bm}[1]{\mbox{\boldmath $#1$}}

\title{Statistical Dynamics of On-line Independent Component Analysis}

\author{Gleb Basalyga {\thanks{E-mail: basalyga@cs.man.ac.uk}} , Magnus Rattray {\thanks{E-mail:
magnus@cs.man.ac.uk}}
\\
\\ {Department of Computer Science, University of Manchester}
\\        {Manchester, M13 9PL, UK}
\\
\\{\small Preliminary version of paper to appear}
\\ {\small in Journal
of Machine Learning Research special issue on ICA} \\ \\}

\begin{document}
\maketitle
\begin{abstract}
The learning dynamics of on-line independent component analysis is
analysed in the limit of large data dimension. We study a simple
Hebbian learning algorithm that can be used to separate out a
small number of non-Gaussian components from a high-dimensional
data set. The de-mixing matrix parameters are confined to a
Stiefel manifold of tall, orthogonal matrices and we introduce a
natural gradient variant of the algorithm which is appropriate to
learning on this manifold. For large input dimension the parameter
trajectory of both algorithms passes through a sequence of
unstable fixed points, each described by a diffusion process in a
polynomial potential. Choosing the learning rate too large
increases the escape time from each of these fixed points,
effectively trapping the learning in a sub-optimal state.  In
order to avoid these trapping states a very low learning rate must
be chosen during the learning transient, resulting in learning
time-scales of $O(N^2)$ or $O(N^3)$ iterations where $N$ is the
data dimension. Escape from each sub-optimal state results in a
sequence of symmetry breaking events as the algorithm learns each
source in turn. This is in marked contrast to the learning
dynamics displayed by related on-line learning algorithms for
multilayer neural networks and principal component analysis.
Although the natural gradient variant of the algorithm has nice
asymptotic convergence properties, it has an equivalent transient
dynamics to the standard Hebbian algorithm.
\end{abstract}

\section{Introduction}

On-line learning algorithms are often used for dealing with very
large data sets or in dynamic situations in which data is changing
according to a non-stationary process. Independent component
analysis (ICA) is often applied under one or both of these
conditions and a number of on-line ICA algorithms have been
developed \citep[see, e.g.][]{hyva99}. In on-line learning the model parameters are updated
after the presentation of each training example. Although there is
good understanding of the asymptotic properties of on-line
learning, much of the learning takes place far from the asymptotic
regime and here a theoretical understanding of the process is
often poor. Training methods are typically based on experimental
observations in the absence of a successful theoretical analysis
of on-line learning in this regime. The efficiency of on--line
training is very sensitive to the choice of training parameters
such as the learning rate and this dependence can slow down learning
and influence the ability of learning to converge successfully to
desired states. A deeper theoretical understanding of the on-line
learning process is needed to provide reliable methods for
setting the parameters and optimising the training process.

Most classical theoretical results on on-line learning are from
stochastic approximation theory~\citep{kushner78}. For a review of
recent advances and modern theoretical approaches, see e.g.
\citet{saad98}. Theories describing on-line learning have mainly
been developed in two different asymptotic regimes. Most work has
been done in the limit of the long times, in which case the model
parameters are close to a stable fixed point of the learning
dynamics. Here one can work out the asymptotic distribution of the
parameters for constant learning rate or study their convergence
to a fixed point as the learning rate is reduced according to some
annealing schedule. The conditions under which the parameters
converge at an optimal rate are quite well
understood~\citep{white89}. Work has also been carried out in the
limit of small learning rates, in which case the dynamics can be
shown to follow the mean gradient or flow
globally~\citep{kushner78}. Unifying these two strands to some
extent one can study the long-time global behaviour under an
annealed learning schedule~\citep{kushner87}.

It is unclear how much practical relevance the classical
asymptotic limits have, since often in practical applications
learning is not asymptotic in the learning times or in the
learning rate. In this work we pursue a different type of
asymptotic analysis by considering the limit of large system size.
This limit has been studied extensively by researchers applying
statistical mechanics methods to the study of learning
systems~\citep{engel01}. For example, the dynamics of gradient
decent in multilayer perceptrons (MLPs) \citep{biehl95,saad95a}
and the dynamics of Sanger's principal component analysis (PCA)
algorithm \citep{biehl94,biehl98} have been studied in this limit.
In these examples the dynamics displays interesting and
non-trivial transient behaviour with unstable, sub-optimal fixed
points appearing due to a symmetry in the parameter space of the
models. The dynamics of natural gradient learning in MLPs has been
studied using these techniques, showing that the transient fixed
point becomes less serious and transient learning performance is
improved compared to standard online gradient
descent~\citep{rattray98,rattray99}. Since a natural gradient
algorithm has been shown to provide an asymptotically efficient
on-line learning algorithm for ICA \citep{amari96,amari98} we are
interested in whether the statistical mechanics approach developed
here can help understand its transient performance.

We have recently developed a novel theoretical framework for
studying the dynamics of on-line ICA in the limit of large data
dimension~\citep{rattray02,nips02-LT19}. We study a Hebbian
learning rule, which is a simple and popular algorithm for on-line
ICA with nice stability conditions~\citep{hyva98} and is closely
related to a popular fixed-point batch algorithm~\citep{hyva97}.
The algorithm is particularly amenable to analysis in the limit of
large input dimension because it can be used to extract a small
number of independent components from  high-dimensional data. We
will see later that this allows the dynamics to be represented by
a relatively small number of variables when the data dimension
becomes large. As well as studying the standard version of the
algorithm we also develop a natural gradient variant and study its
dynamics. Because the parameter space is constrained to the
Stiefel manifold of orthogonal rectangular matrices, the standard
equivariant or natural gradient ICA algorithms due to
\citet{cardoso96} and \citet{amari96} are not appropriate here.
Instead it is possible to use the ideas developed by
\citet{edelman99geometry} in order to construct an algorithm which
follows the gradient on a Stiefel manifold. Our variant uses the
gradient defined by~\citet{amari1999} but includes the
orthogonalisation term from~\citet{hyva98} in order to keep the
model parameters on the Stiefel manifold.

In order to obtain a tractable ICA model, we consider an idealised
data set in which a small number of non-Gaussian sources are mixed
into a large number of Gaussian sources \citep{rattray02}. By
using the methods of statistical mechanics,  we provide a solution
to the dynamics of Hebbian ICA in the limit of large input
dimension. This generalises on previous results
\citep{rattray02,nips02-LT19}, which were limited to the simplest
single source case (except for the late time asymptotics, which
were solved for the general case). We find that the transient
dynamics of Hebbian ICA can be described as a stochastic process
in which the system moves through a sequence of metastable fixed
points, each of which can be described as a multi-dimensional
diffusion in polynomial potential. By solving the dynamics of the
multi-source case we can characterise the symmetry breaking
process which is critical to performance of the learning process.

It is interesting to observe that the dynamics of ICA is very
different from the dynamics of learning in MLPs and in PCA studied
previously~\citep{saad95a,biehl98}. In these cases the dynamics
was observed to be ``self-averaging'' so that it followed a smooth
trajectory in the limit of large data dimension $N$ and the
learning happened on an $O(N)$ time-scale. The ICA dynamics
displays significant fluctuations even in this limit and learning
occurs on a much slower time-scale, typically requiring of the
order of $N^2$ or $N^3$ iterations depending on the details. We
find that the natural gradient variant of the Hebbian algorithm
has very nice asymptotic convergence properties with uniform
convergence in the case of equal source statistics. However, the
algorithm is shown to have equivalent transient performance to the
standard algorithm.

This paper is organised as follows. The data model is described in
Section \ref{sectionII}. The on-line Hebbian ICA algorithm is
introduced in Section \ref{sectionIII}. In
Section~\ref{sec_largeN} we introduce the macroscopic variables
which provide a compact description of the dynamics for large $N$.
In Section \ref{sectionIV} we show that the learning dynamics near
a sub-optimal fixed point close to the initial conditions can be
considered as a diffusion process. The transient dynamics through
a sequence of metastable states is analysed in Section
\ref{sec_transient}. In Section \ref{sec_nat} we introduce the
natural gradient version of Hebbian ICA algorithm and study its
dynamics. General conclusions are made in Section
\ref{sec_conclusion}.

\section{Data Model}
\label{sectionII}

We consider the following idealised linear data model introduced
in  \citet{rattray02}. The $N$-dimensional data $\bm x$ is
generated from a noiseless linear mixture of a small number $M$ of
non-Gaussian sources $\bm s$
 and a large
number $N-M$ of uncorrelated Gaussian components, $\bm n\sim{\cal
N}(\bm 0,\bm I_{N-M})$,
\begin{equation}
 \bm x =
 \bm A \left[\begin{array}{c} \bm s \\ \bm n
 \end{array}\right]=
\bm A_s \bm s + \bm A_n \bm n , \label{eqn_x}
\end{equation}
where $\bm A = [\bm A_s \ \bm A_n]$ is the $N\times N$ mixing
matrix, $\bm I_{N}$ denotes an $N\times N$ identity matrix and
 ${\cal N}(\bm a,\bm \Sigma)$ denotes a Gaussian distribution with
mean $\bm a$ and covariance matrix $\bm \Sigma$. Without loss of
generality it can be assumed that the sources each have unit
variance. In order to apply the Hebbian ICA algorithm
 we assume also
that the data is already sphered, i.e. the data has zero mean and
an identity covariance matrix.  This can be achieved for on-line
learning by an adaptive sphering algorithm, such as the one
introduced by~\cite{cardoso96}. These model assumptions lead to
the constraint that $\bm A$ must be an orthogonal matrix, e.g.
\begin{eqnarray} \left[\bm A_s \; \bm
A_n\right]\left[\begin{array}{c} \bm A_s^\tr \\ \bm A_n^\tr
\end{array}\right] & = & \bm A_s \bm A_s^\tr + \bm A_n \bm A_n^\tr = \bm
I \ , \label{eqn_Ac1} \\ \left[\begin{array}{c} \bm A_s^\tr \\ \bm
A_n^\tr \end{array}\right]\left[\bm A_s \; \bm A_n\right] & = &
\left[\begin{array}{c c} \bm A_s^\tr \bm A_s & \bm A_s^\tr \bm A_n \\
\bm A_n^\tr \bm A_s & \bm A_n^\tr \bm A_n \end{array}\right] =
\left[\begin{array}{c c} \bm I & \bm 0 \\ \bm 0 & \bm I
\end{array}\right]. \label{eqn_Ac2}
\end{eqnarray}

\section{On-line Hebbian  ICA Learning Rule}
\label{sectionIII}

The goal of ICA is to find  the de-mixing matrix $\bm W$ such that
the projections,
\begin{equation}\label{projections}
\bm y \equiv \bm W^\tr \bm x \ ,
\end{equation}
 will coincide with
the non-Gaussian sources $\bm s$ up to scaling and permutations.
The best possible solution is one in which the $K$ projections
will learn as many as possible of the $M$ non-Gaussian sources.
Note that specialisation of each projection to a particular source
mostly depends on the details of the initial conditions.

We consider a simple Hebbian learning rule \citep{hyva98}, which
extracts non-Gaussian sources from the data by maximising some
measure of non-Gaussianity of the projections. The change of the
$N\times K$ de-mixing matrix $\bm W$ at the each learning step is
given by,
\begin{equation}
   \bm \Delta \bm W = \eta \, \bm x
\bm \phi(\bm y)^\tr \bm \sigma  + \alpha \bm W(\bm I - \bm W^\tr\bm W)
\ . \label{eqn_dw}
\end{equation}
 Here, $\eta$ is the
learning rate and $\bm \sigma$ is a $K\times K$ diagonal matrix
with elements
\begin{equation}
   \sigma_{ii} = \mbox{sign}\left(\mbox{E}_{s_i}[s_i \phi(s_i) -
  \phi'(s_i)]\right) ,
\label{stab}
\end{equation}
which ensures stability of the correct solution as $y_i
\rightarrow s_i$ \citep{hyva98}. The first term on the right of
(\ref{eqn_dw}) maximises some measure of non-Gaussianity of the
projections. The second term provides orthogonalisation of the
de-mixing matrix. The choice of learning rate $\eta$ greatly
influences the performance of this algorithm. Choosing too large a
learning rate results in slow and inefficient learning but choosing
too high a value may result in the learning dynamics becoming
trapped in a poor solution, as we will see later. The learning
dynamics is less sensitive to the choice of the parameter $\alpha$
and we set $\alpha=0.5$ in simulations. The function $\bm \phi(\bm
y)=[\phi(y_{i})]$ is some smooth non-linear function which is
applied to every component of the vector $\bm y$. An even
non-linearity, e.g. $\phi( y)= y^2$, is usually used to detect
asymmetric non-Gaussian signals, while an odd non-linearity, e.g.
$\phi( y)= y^3$ or $\phi( y)=\tanh( y)$, is used to extract
symmetric non-Gaussian signals.

\section{Learning Dynamics for Large Data Dimension}
\label{sec_largeN}

In the case of high-dimensional data (when $N$ becomes very big)
it is difficult to analyse the dynamics of the $N\times K$
de-mixing matrix. In order to provide a compact description of the
system dynamics in the limit $N\rightarrow \infty$, we introduce
new macroscopic variables
\begin{eqnarray}
    \bm R\equiv \bm W^\tr\bm A_s  \quad \mbox{ and } \quad
 \bm Q \equiv \bm
W^\tr\bm W\ ,\label{overlaps}
\end{eqnarray}
the dimension of which are $K\times M$ and $K\times K$
respectively. These overlap matrices contain all necessary
information about the relationship between the projections $\bm y$
and the sources $\bm s$. Therefore the system can  be described by
a small number of macroscopic quantities in the limit of large $N$
as long as $K$ and $M$ remain small. We will usually order the
indices retrospectively so that the dynamics approaches the
optimal solution with $R_{ij}=\delta_{ij}$ (assuming orthogonal
$\bm W$ such that $\bm Q=\bm I$). However, it should be remembered
that there are equivalent optima related to this solution by a
permutation of indices or changes in sign of the components of
$\bm R$. In order to learn successfully the algorithm has to break
 symmetry and specialise to a particular solution.

Using Equation~(\ref{eqn_Ac1}) one can show that,
\begin{eqnarray}
    \bm y =  \bm W^\tr(\bm A_s \bm s + \bm A_n \bm n)  =
 \bm R\bm s + \bm z \ ,\label{eqn_y}
\end{eqnarray}
where $\bm z\sim{\cal N}(\bm 0,\bm C)$ and $\bm C = \bm Q - \bm R
\bm R^\tr$. In order to analyse the dynamics we will have to
compute expectations with respect to the distribution of $\bm
y$. The covariance matrix $\bm C$ is symmetric and positive
definite so that we can always write $\bm z$ in the form
   $\bm z=\bm L \bm \mu$ where $\bm \mu\sim{\cal
N}(\bm 0, \bm I)$ are Gaussian variables and the matrix $\bm L$
 can be found by special decomposition. A particularly useful
decomposition for our purposes is the Cholesky decomposition
\citep[see, e.g.][]{crecipes} in which case $\bm C =\bm L \bm
L^\tr$ where $\bm L$ is lower-triangular with non-zero components
satisfying the following recurrence relations
\begin{equation}
L_{kk}=\sqrt{C_{kk}-\sum_{j=1}^{k-1} L_{kj}^2}
\label{chol1}
\end{equation}
and for the $k$-th row of $\bm L$ we have
\begin{equation}
L_{ki}= \frac{C_{ki}-\sum_{j=1}^{i-1} L_{ij} L_{kj}}{L_{ii}} \ ,
\label{chol2}
\end{equation}
 for  $i=1,2,...,k-1$.

From Equation~(\ref{eqn_dw}) we can calculate the changes in $\bm
R$ and $\bm Q$  after a single learning step,
\begin{eqnarray}
    \Delta \bm R & = & \eta \bm \sigma \bm\phi(\bm y)\bm s^\tr + \alpha (\bm I-\bm Q)\bm R \ , \label{dr} \\
    \Delta \bm Q & = & \eta \bm \sigma (\bm I+\alpha(\bm I - \bm
    Q))\bm\phi(\bm y)\bm y^\tr + \alpha^2(\bm I - \bm
    Q)^2\bm Q \nonumber \\
  &  +& \eta\bm\sigma\bm y\bm\phi(\bm
    y)^\tr(\bm I + \alpha(\bm I - \bm Q))  + 2\alpha(\bm I - \bm Q)\bm Q \nonumber
 \\ & + & \eta^2\bm\phi(\bm y)\bm x^\tr\bm x\bm\phi(\bm y)^\tr \
    ,
    \label{dq}
\end{eqnarray}
where we used the constraint in Equation~(\ref{eqn_Ac2}) to set
$\bm x^\tr \bm A_s = \bm s^\tr$.

The dynamics is not very sensitive to the exact value of $\alpha$
as long as $\alpha\gg\eta$. We will see later that the learning
rate must be chosen very small for large $N$ so that typically we
will have this situation in practice. As $\alpha$ increases
relative to $\eta$, $\bm Q$ approaches $\bm I$ since the
orthogonalisation term in Equation~(\ref{eqn_dw}) dominates. If
one defines $\bm Q-\bm I\equiv \bm q/\alpha$ and sets $\alpha$
large relative to $\eta$ then the fixed point of
Equation~(\ref{dq}) to leading order is,
\begin{equation}
    \bm q = \mbox{$\frac{1}{2}$} \left[\eta \bm\sigma \left(\bm\phi(\bm y)\bm
    y^\tr + \bm y \bm\phi(\bm y)^\tr\right) + \eta^2 N \bm\phi(\bm
    y)\bm\phi(\bm y)^\tr\right] \ ,
   \label{qq}
\end{equation}
where we have dropped terms lower than $O(\eta^2 N)$ and
$O(\eta)$. Substituting Equation (\ref{qq}) into
Equation~(\ref{dr}) leads to the following update equation for
$\bm R$,
\begin{eqnarray}
 \Delta \bm R = \eta  \bm\sigma\left[ \bm\phi(\bm
y)\bm s^\tr - \mbox{$\frac{1}{2}$} \left(\bm\phi(\bm y)\bm y^\tr +
\bm y\bm\phi(\bm y)^\tr\right)\bm R\right] - \mbox{$\frac{1}{2}$}\
\eta^2 N \bm\phi(\bm y)\bm\phi(\bm y)^\tr \bm R\ .
\label{eqn_dR}
\end{eqnarray}
This simplification procedure is an example of adiabatic
elimination of fast variables \citep[see, for
example,][]{gardiner}. In a more rigorous treatment one should
consider the mean and covariance of $\Delta \bm R$ and $\Delta \bm
Q$ using the appropriate large $N$ scalings which are described in
the next sections. One finds that fluctuations in $\bm Q$ are
negligible in the limit of large $N$ and therefore the dynamics of
$\bm Q$ can be described by a differential equation in this  limit
 with stable fixed point given by Equation~(\ref{qq}).

\begin{figure*}[t!]
    \setlength{\unitlength}{0.9cm} \begin{center}
    \begin{picture}(9,4) \epsfysize = 5.5cm
    \put(-3.2,-2.2){\epsfbox[50 0 550 600]{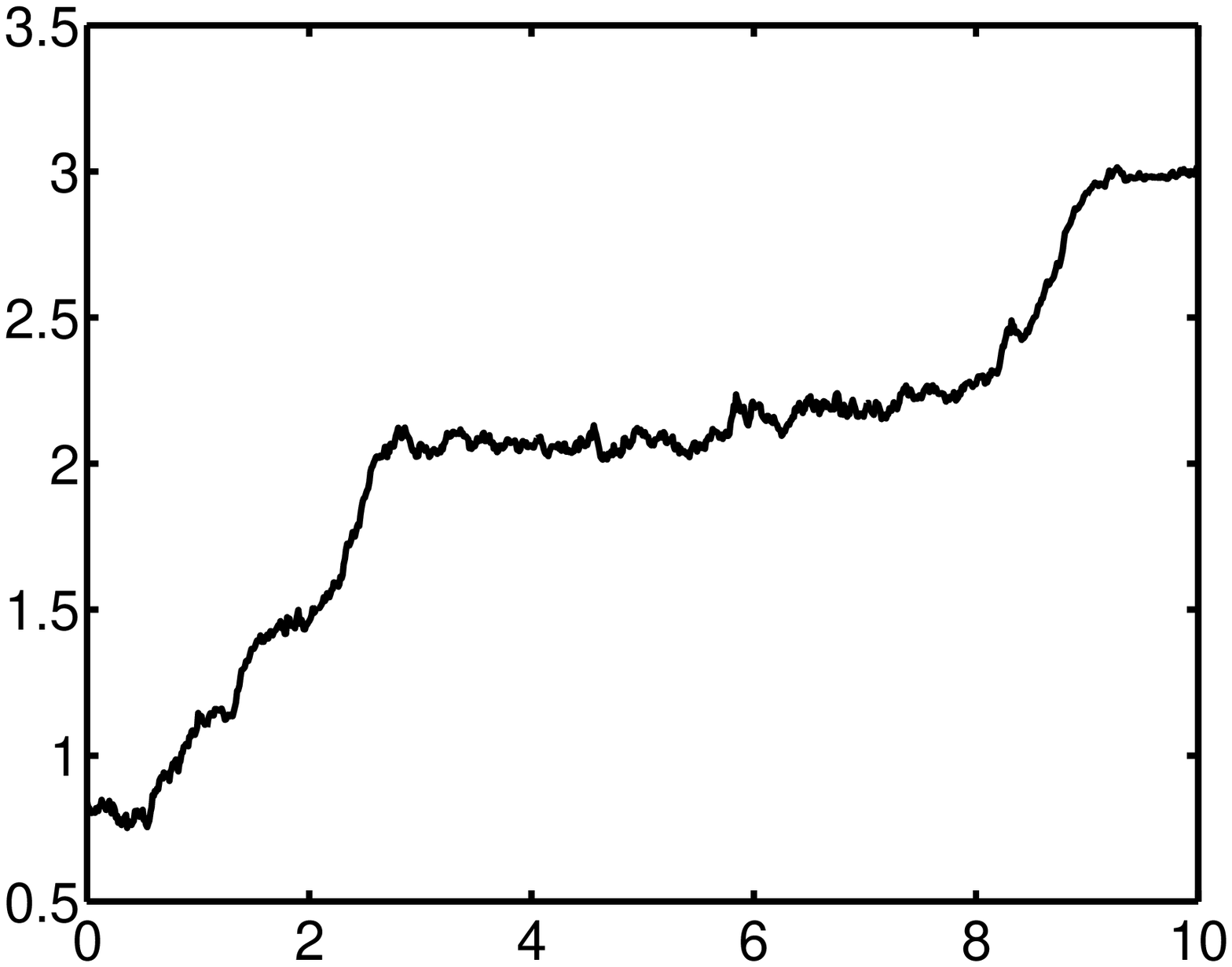}}
    \put(-2.3,3){\mbox{(a)}}
    \put(-1.6,4){\mbox{$\eta=3.5\times 10^{-4}$}}
    \put(-4.8,2.5){\mbox{$\sum_{ij}|R_{ij}|$}}
    \put(-0.75,-0.6){\mbox{$t/N^2$}}
            \epsfysize =5.5cm \put(2.3,-2.2){\epsfbox[50 0 550 600]{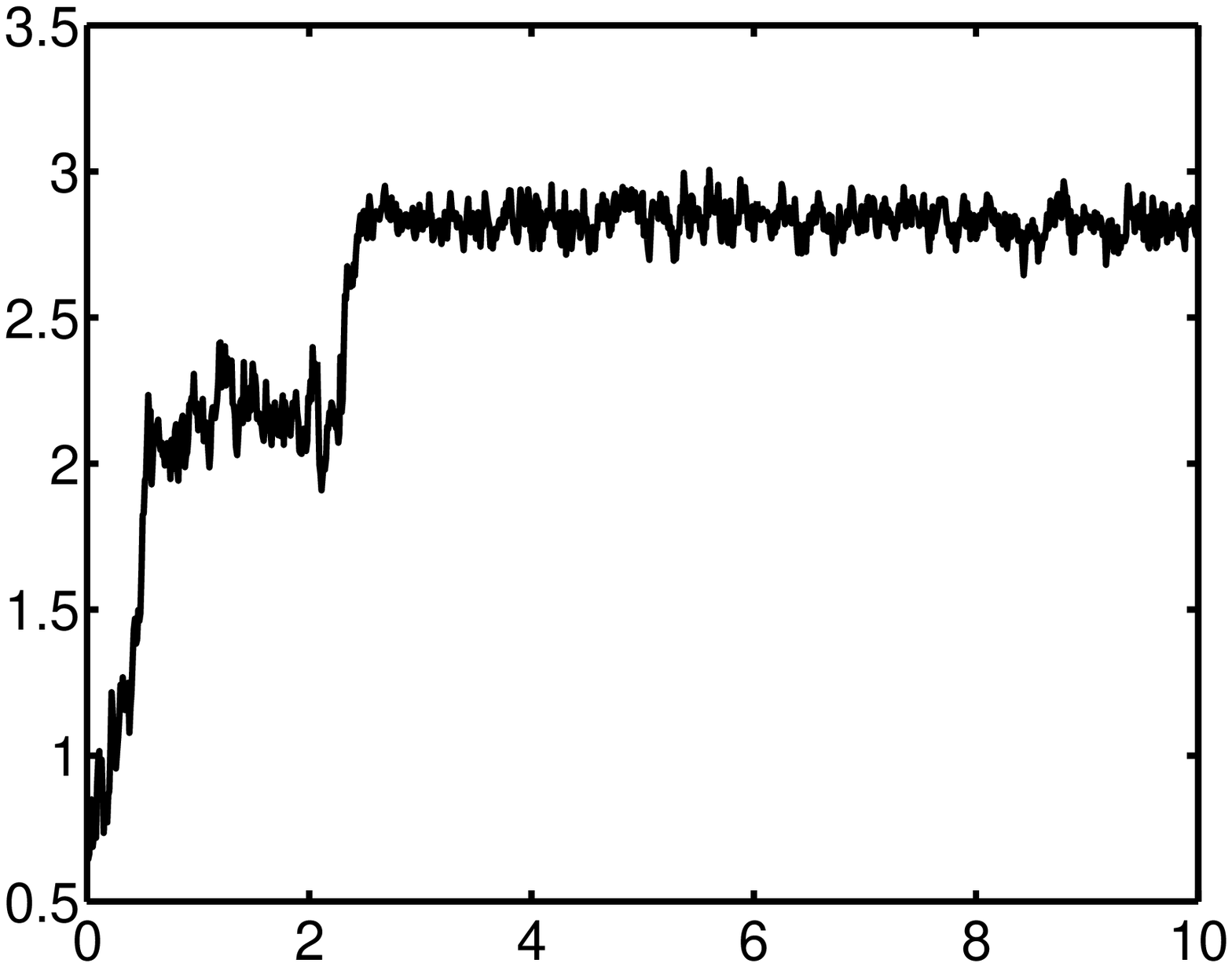}}
        \put(4.1,4){\mbox{$\eta=0.0025$}}
         \put(3.3,3){\mbox{(b)}}
           \put(4.71,-0.6) {\mbox{$t/N^2$}}
       \epsfysize =5.5cm \put(7.8,-2.2){\epsfbox[50 0 550 600]{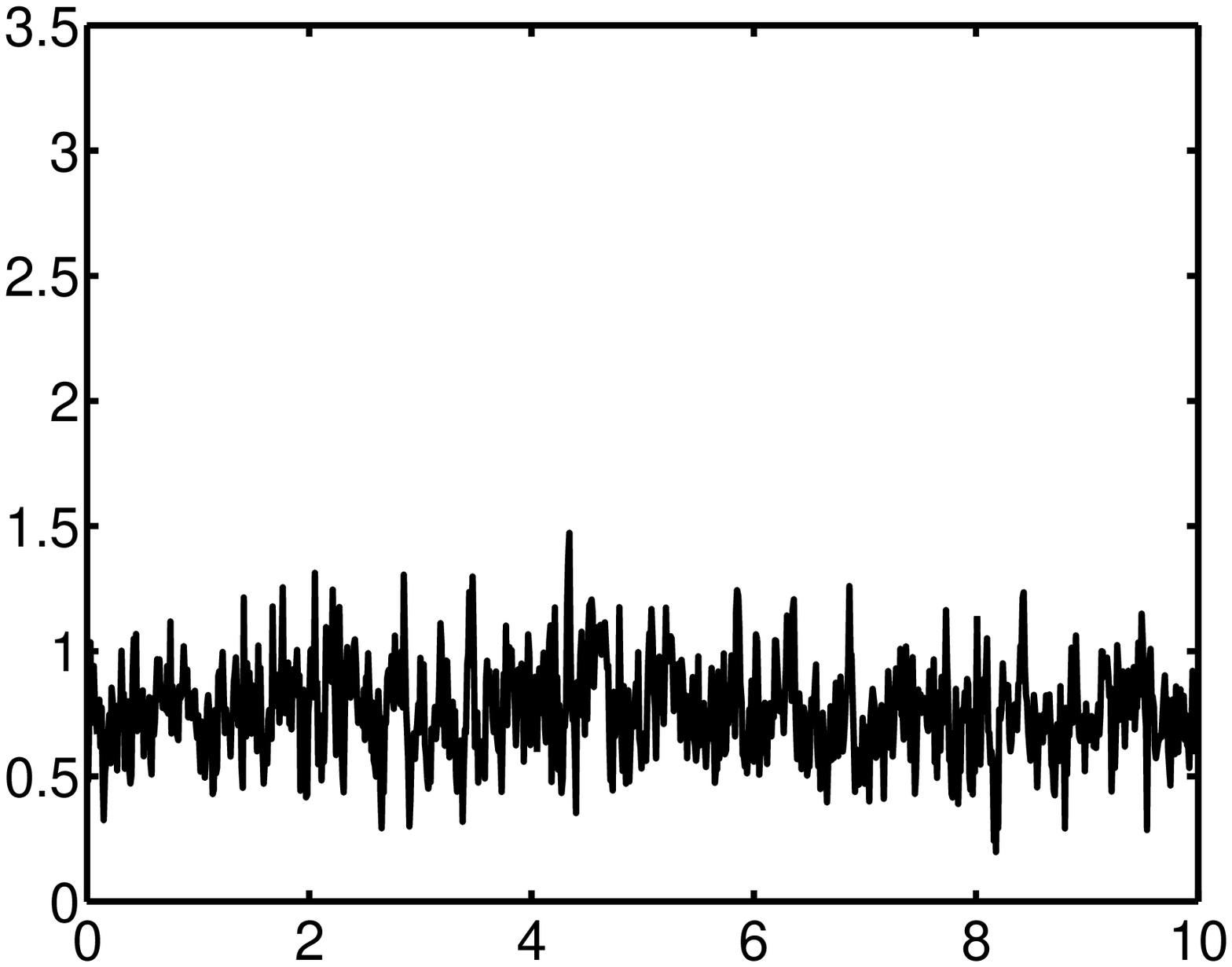}}
        \put(9.8,4){\mbox{$\eta=0.005$}}
         \put(8.8,3){\mbox{(c)}}
          \put(10.2,-0.6) {\mbox{$t/N^2$}}
         \end{picture} \end{center}
        \caption{\label{fig_general} Typical learning dynamics of the Hebbian ICA algorithm for different learning rates.
        We used the non-linearity $\phi(y)=y^2$ to extract
        three asymmetrical binary sources with
    skewness $\kappa_3=1.5$ from 100-dimensional data ($K=M=3$, $N=100$).
 Each picture shows the quantity $\sum_{ij}|R_{ij}|$ as it changes
    over time. For the smallest learning rate ($\eta=3.5\times
    10^{-4}$) the dynamics looks
 relatively smooth but it takes time to learn all three sources and
    the dynamics is localised at a sub-optimal metastable state near
    $\sum_{ij}|R_{ij}| = 2$ for a significant period of time. With a larger
 learning rate ($\eta=0.0025$) the dynamics appears more stochastic
    and the system is again localised in the same metastable
    state. For the largest learning rate ($\eta=0.005$) the
    fluctuations are so strong that the system remains trapped in a
    sub-optimal state close to the initial conditions for the entire
    simulation time.
     }
\end{figure*}

In Figure~\ref{fig_general} we show some typical dynamical
trajectories. We observe the following types of fixed points in
the $\bm R$-dynamics:
\begin{itemize}
  \item Optimal fixed points (see Figure \ref{fig_general} (a) and (b))

Asymptotically, when $y_i\rightarrow s_i$,  the optimal solution
is given by $R^*_{ij} =\delta_{ij}$, which is a fixed point of
Equation (\ref{eqn_dR}) as $\eta\rightarrow 0$. Note that all
other possible solutions can be obtained by a trivial permutation
of indices and/or changes in sign. Asymptotically the learning
rate should be annealed in order to approach this fixed point at
an optimal rate. For a detailed account of the asymptotic dynamics
of the Hebbian ICA algorithm under an annealed learning rate, see
\cite{rattray02}.

\item Sub-optimal fixed points causing trapping near the  initial
  conditions (see Figure \ref{fig_general} (c))

   Due to the $O(\eta^2)$ fluctuation
term in Equation (\ref{eqn_dR}), the  algorithm has a special
class of sub-optimal fixed points near $\bm R=\bm 0$ which causes
the presence of a stochastic trapping state near the initial
conditions. We will discuss this situation in detail in the next
Section.

 \item Transient fixed points (see Figure \ref{fig_general} (a) and (b))

When $T<M$ non-Gaussian sources have been learned, the dynamics
can become localised in metastable states somewhere between the
initial conditions and the final, optimal fixed point. In this
case the fixed points of Equation~(\ref{eqn_dR}) are
\begin{equation}\label{tfp}
R^*_{ij} =\delta_{ij} \  \mbox{I}\left[i \leq T\right],
\end{equation}
where
\begin{equation}\label{predicate}
\mbox{I}\left[\mbox{predicate}\right]=\left\{\begin{array}{l} 1 \
\mbox{ if predicate is true}, \\
        0  \ \ \mbox{if predicate is false}. \end{array}\right .
\end{equation}
Again, similar fixed points can be obtained from (\ref{tfp}) by a
simple permutation of indices. In this case the system has to
leave such metastable states in order to complete learning. We
will analyse the dynamics near these metastable states in Section
\ref{sec_transient}.
\end{itemize}

\section{Stochastic Trapping State near the Initial Conditions}
\label{sectionIV}

In similar studies of on-line learning, macroscopic quantities
like the overlap $\bm R$ usually have a ``self-averaging''
property such that the variance of these macroscopic quantities
tends to zero in the limit $N\rightarrow \infty$ ~\citep[see
e.g.][]{saad95a,biehl98}. A random and uncorrelated choice for
$\bm A$ and the initial entries of de-mixing matrix $\bm W$
leads us to expect $\bm R = O(N^{-1/2})$ initially. Larger initial
values of $\bm R$ could only be obtained with some prior knowledge
of the mixing matrix which we will not suppose. In this case one
can no longer assume that fluctuations are negligible as
$N\rightarrow \infty$. Moreover, as we will see below, the mean
and variance of the change in $\bm R$ at each iteration are of the
same order. That means that the overlap $\bm R$ does not
self-average and the fluctuations have to be considered even in
the limit. Therefore, it is more natural to model the on-line
learning dynamics near the initial conditions as a diffusion
process ~\citep[see, for example,][]{gardiner,kampen}. In order to
establish a clear picture of the dynamics we have to choose an
appropriate scaling for macroscopic quantities and learning
parameters.
 In the following discussion we set $\bm r\equiv
\bm R\sqrt{N}$ where $\bm r$ is assumed to be an $O(1)$
quantity.

\subsection{Diffusion in a Potential}

For a diffusion process the probability density $p(\bm r,t)$ of
 a random variable $\bm r = \left[r_{ij}\right]$ at time $t$ obeys the Fokker-Planck equation
\begin{eqnarray}\label{fpeq}
  \frac{\partial p(\bm r,t)}{\partial t}=-\sum_{ij} \frac{\partial }{\partial
  r_{ij}} \left (A_{ij} \ p(\bm r,t)\right)
  +\frac{1}{2}\sum_{ijkl}\frac{\partial^2 }{\partial
  r_{ij}\partial r_{kl}}\left (B_{ijkl} \ p(\bm r,t)\right) \ ,
\end{eqnarray}
where the coefficients $A_{ij}$, which are called
``drift'' coefficients, represent the expectation of the change of
variable $r_{ij}$ with respect to the stochastic process in
question. They can be written as
 \begin{equation}\label{A}
  A_{ij}=\mbox{E}[\Delta r_{ij}]=-\frac{\partial U(\bm
  r)}{\partial  r_{ij}} N^{-p} \ ,
\end{equation}
where $N$ and $p$ are the input dimension and the scaling order
for our system and $U(\bm r)$ is some differentiable function of
$\bm r$. This function is analogous to the potential function for the
case of a particle undergoing a diffusion in a potential. The
coefficients $B_{ijkl}$ are the covariance of the change of variable $r_{ij}$,
\begin{equation}\label{B}
  B_{ijkl}=\mbox{Cov}[\Delta
    r_{ij}, \Delta r_{kl}]=\mbox{E}[\Delta
r_{ij}\Delta r_{kl}] -\mbox{E}[\Delta r_{ij}] \mbox{E}[\Delta
r_{kl}] =D_{ijkl} N^{-p} ,
\end{equation}
where  $D_{ijkl}$ is called the ``diffusion term''. Usually this would
be a matrix but notice that in our case the dynamical variables are
in a matrix and therefore the diffusion term has four
indices. However, we can think of each pair as a single index in a
vectorised system. If we do so then for our case the diffusion term can be
considered a diagonal matrix of magnitude $D$,
\begin{equation}\label{D}
 D_{ijkl}=D \ \delta_{ik}  \delta_{jl} ,
\end{equation}
where $D$ is called the ``diffusion coefficient''. It is typical for
the diffusion process that the mean and covariance of the random
variable are of the same order $\sim O(N^{-p})$. This means that,
for large $N$, the system is equivalent to diffusion in
the potential $U(\bm r)$ with a characteristic time-scale $(\delta
t)^{-1} = N^{p}$.

\begin{figure*}[t]
    \setlength{\unitlength}{0.8cm} \begin{center}
    \begin{picture}(9,4) \epsfysize = 12.4cm
    \put(-2.5,-0.9){
\epsfbox[50 0 550 600]{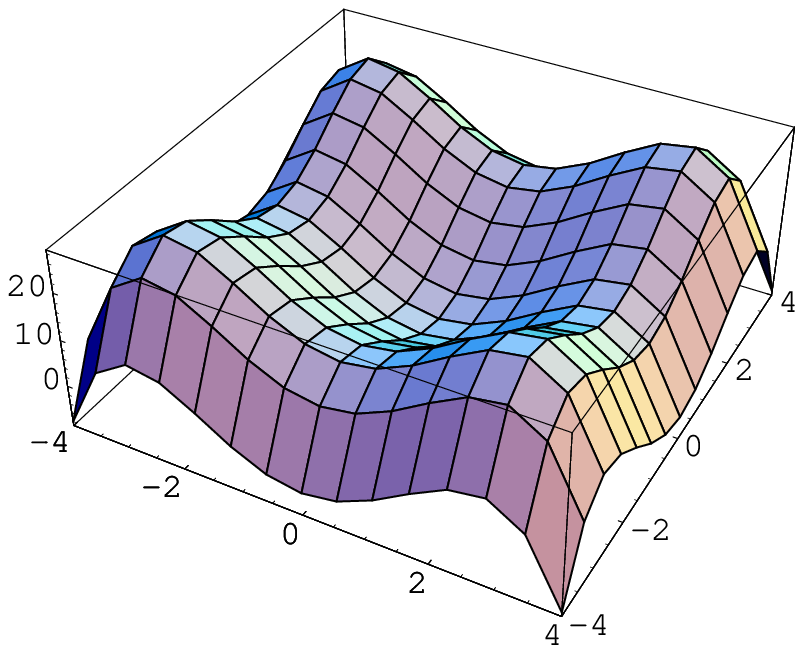}}
    \put(+0,4.3){\mbox{$\phi(y)$ odd, \ $\kappa_4\neq 0$}}
    \put(-2,3.3){\mbox{$U(r_{11},r_{12})$}}
    \put(0.4,-0.4){\mbox{$r_{11}$}}
        \put(4.2,0.5){\mbox{$r_{12}$}}
        \epsfysize = 13cm \put(4,-1){\epsfbox[50 0 550 600]{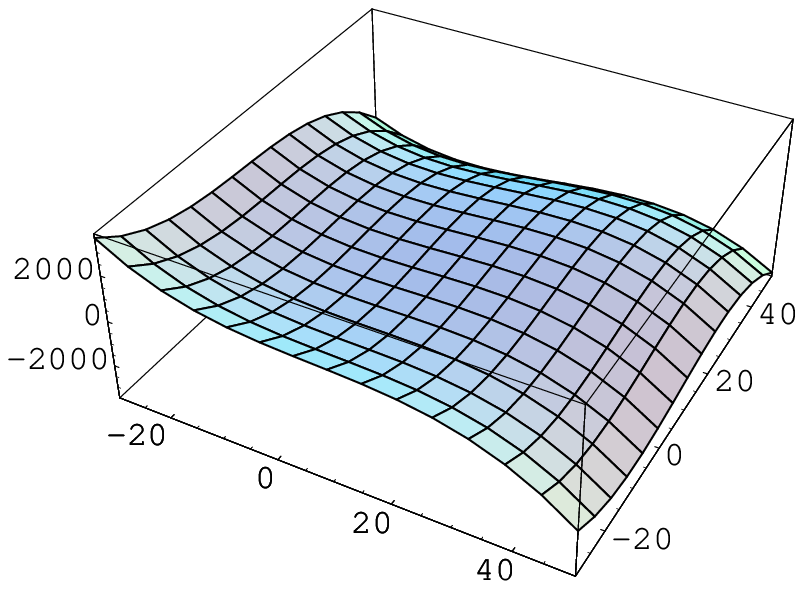}}
        \put(6.5,4.3){\mbox{$\phi(y)$ even, \ $\kappa_3\neq 0$}}
         \put(5.1 ,3.4){\mbox{$U(r_{11},r_{12})$}}
        \put(6.8,-0.3){\mbox{$r_{11}$}}
        \put(10.8,0.15){\mbox{$r_{12}$}}
         \end{picture} \end{center}
        \caption{Close to the initial conditions the learning dynamics is
        equivalent to diffusion in a polynomial potential.  For symmetrical source distributions
         with non-zero kurtosis $\kappa_4$ we should use an odd non-linearity in which
        case the potential is quartic, as shown on the left (for  $K=1$, $M=2$). For
        asymmetrical source distributions with  skewness $\kappa_3$  we can use an even
        non-linearity in which case the potential is cubic, as shown
        on the right. Escaping over the minimal barriers results in symmetry breaking,
         i.e. specialisation to a particular source.}  \label{fig_potential}
\end{figure*}

In our case the potential $U(\bm r)$ has a minimum at $r_{ij}=0$
surrounded by potential barriers of differing heights. Examples
are shown in Figure~\ref{fig_potential} for an odd and even
non-linearity on the left and right respectively. The escape time
(the mean first passage time) from the minimum of the potential at
$\bm r=\bm 0$ is mainly determined by the effective size of the
minimal potential barrier $\Delta U$ ~\citep[see, for
example,][]{gardiner,kampen},
\begin{equation}\label{escape}
  T_{\mbox{\tiny escape}}=
A \exp\left(\frac{\Delta U}{D}\right)\ ,
\end{equation}
where  prefactor $A$ is proportional to the characteristic
time-scale and  depends on the curvature of the potential. Because
diffusion through a higher potential barrier is exponentially less
likely as the difference between barrier heights increases, the
escape time will effectively be inversely proportional to the
number of escape points (the number of barriers with the same
minimal height $\Delta U$) when the Arrhenius factor $\Delta U/D$
is large.

In the next two sections we will find the form of potential
barrier and estimate the escape time from the trapping state for
the two main classes of non-linear function $\phi(y)$.

\subsection{Odd Non-linearity}

If we need to extract  symmetrical non-Gaussian signals from the
data then we have to use an odd non-linearity, e.g. $\phi(y)=y^3$
or $\phi( y)=\tanh( y)$ are common choices. In this case the
appropriate scaling for the learning rate will be $\eta = \nu
N^{-2}$, where $\nu$ is an $O(1)$ scaled learning rate parameter.
After expanding Equation (\ref{eqn_dR}) near $\bm r=\bm 0$ we
obtain the following expressions for the mean and covariance of
the change in $ \bm r$ at each iteration,
\begin{eqnarray}\label{mean_odd}
    &&\mbox{E}[\Delta r_{ij}] =   \left( -
    \mbox{$\frac{1}{2}$}\langle\phi^2(\mu)\rangle\nu^2 r_{ij} +
    \mbox{$\frac{1}{6}$}\kappa^j_4\langle \phi'''(\mu)
    \rangle \sigma_{ii} \nu \, r_{ij}^3\right)N^{-3} + O(N^{-4}) ,  \\
     &&\mbox{Cov}[\Delta
    r_{ij}, \ \Delta r_{kl}]   =   \langle \phi^2(\mu)\rangle
    \nu^2 \ \delta_{ik}  \delta_{jl} \ N^{-3} + O(N^{-4})  ,
\label{cov_odd}
\end{eqnarray}
where $\kappa^j_4$ is the fourth cumulant of the $j$-th source
distribution (measuring kurtosis) and brackets denote averages
over a Gaussian variable $\bm \mu\sim{\cal N}(\bm 0,\bm I)$. In
this case the system can be described by a Fokker-Planck equation
for large $N$ with a characteristic time-scale $(\delta
t)^{-1}=N^{3}$. To compute the expectations we have made use of
the Cholesky decomposition of $\bm y$ as described in
Equations~(\ref{chol1}) and (\ref{chol2}). This allows us to
remove the dependence of the parameters in the Gaussian averages
by writing $\bm y =\bm L\bm \mu$ where the lower diagonal (i.e.
non-zero) elements of $\bm L$ are found to be,
\[
    L_{ij} = \delta_{ij} -
    \left(\mbox{$\frac{1}{2}$}\delta_{ij}\sum_m^M
    r_{im}^2 + \sum_{m}^M r_{im}r_{jm}\right)N^{-1} + O(N^{-2}) \quad
    \mbox{for} \quad  i\geq j  .
\]

The dynamics is equivalent to diffusion in the following potential
\begin{equation}
    U(\bm r) = \sum^{K}_{i=1}\sum^{M}_{j=1}\left(\mbox{$\frac{1}{4}$}\langle \phi^2(\mu) \rangle \ \nu^2 r_{ij}^2
    - \mbox{$\frac{1}{24}$}\kappa^j_4 \langle \phi'''(\mu)\rangle
    \sigma_{ii}
    \nu \, r_{ij}^4 \right)
\end{equation}
with a diagonal diffusion matrix of magnitude
$D=\langle\phi^2(\mu)\rangle \nu^2$. Notice that the potential is a
sum of contributions $U(\bm r) = \sum_{ij}U(r_{ij})$ which depend only
on a single element in $\bm r$. Since
the diffusion matrix is diagonal, this means that each element of $\bm
r$ undergoes an independent diffusion process equivalent to the
one-dimensional case described by \cite{rattray02}. The
effective size of the potential barriers for each element $r_{ij}$ is given by,
\begin{equation}\label{Afactor}
      \frac{\Delta U(r_{ij})}{D}=\frac{3\langle
    \phi^2(\mu)\rangle\nu}{8|\kappa_4^j \langle\phi'''(\mu)\rangle|} .
\end{equation}
Escape over one such potential barrier results in the
corresponding source $j$ being learned by projection $i$. This
breaks the symmetry of the system as one projection specialises to
a particular source. Once this happens the system can again become
trapped in a metastable state with other sources remaining
unlearned. The dynamics in the neighbourhood of this more general
class of fixed point is described in Section~\ref{sec_transient}.

\begin{figure*}[t!]
    \setlength{\unitlength}{0.9cm} \begin{center}
    \begin{picture}(9,4) \epsfysize = 6cm
    \put(-2,-2.5){\epsfbox[50 0 550 600]{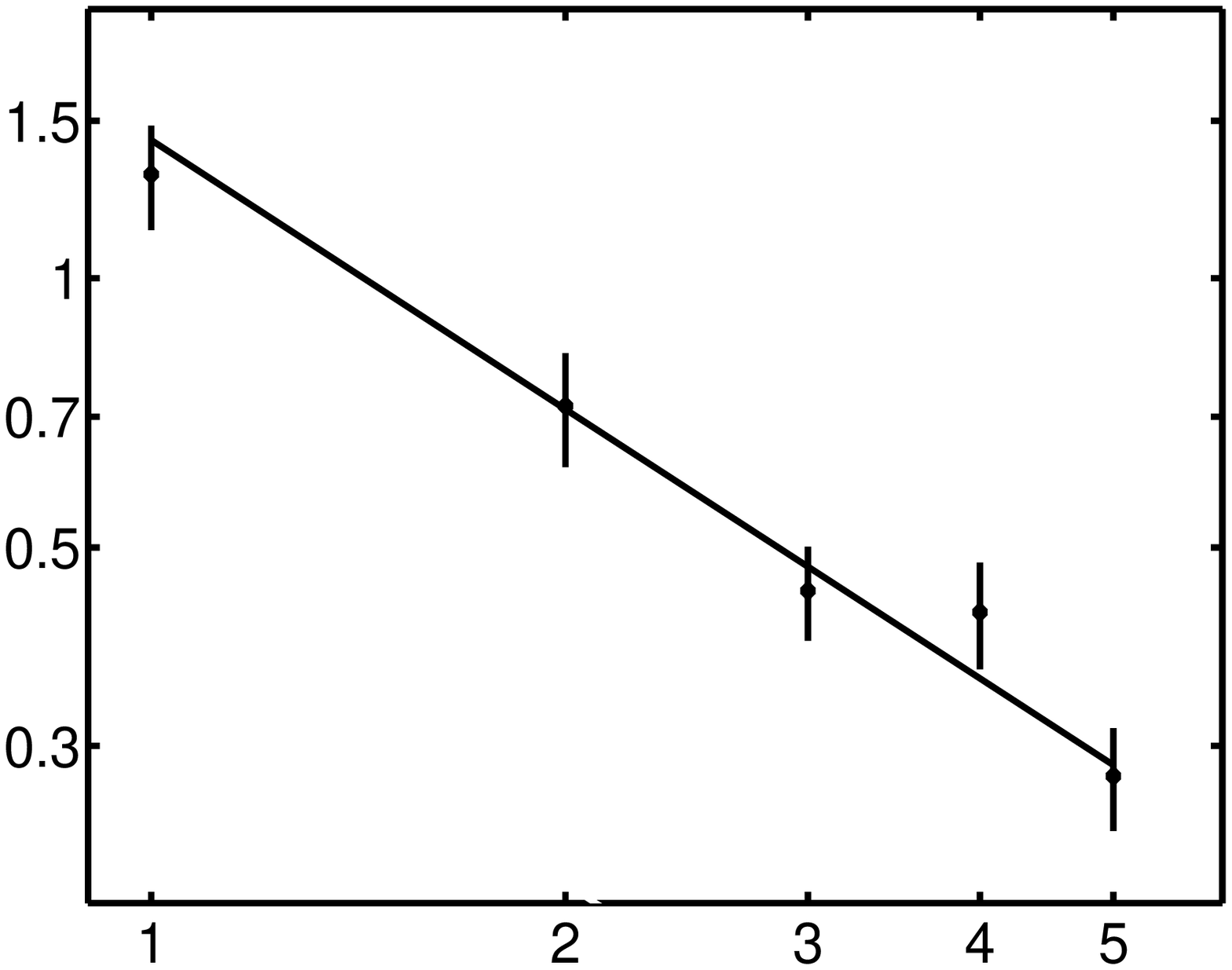}}
    \put(-1.6,4.3){\mbox{$\phi(y)=y^3$,\ $\nu=1$, \ $\kappa_4=-4/9$}}
    \put(-2.8,2.5){\Large{$\frac{T^{\mbox{\tiny odd}}_{\mbox{\tiny escape}}}{N^3}$}}
    \put(0.9,-0.6){\mbox{$M$}}
            \epsfysize =6cm \put(5,-2.5){\epsfbox[50 0 550 600]{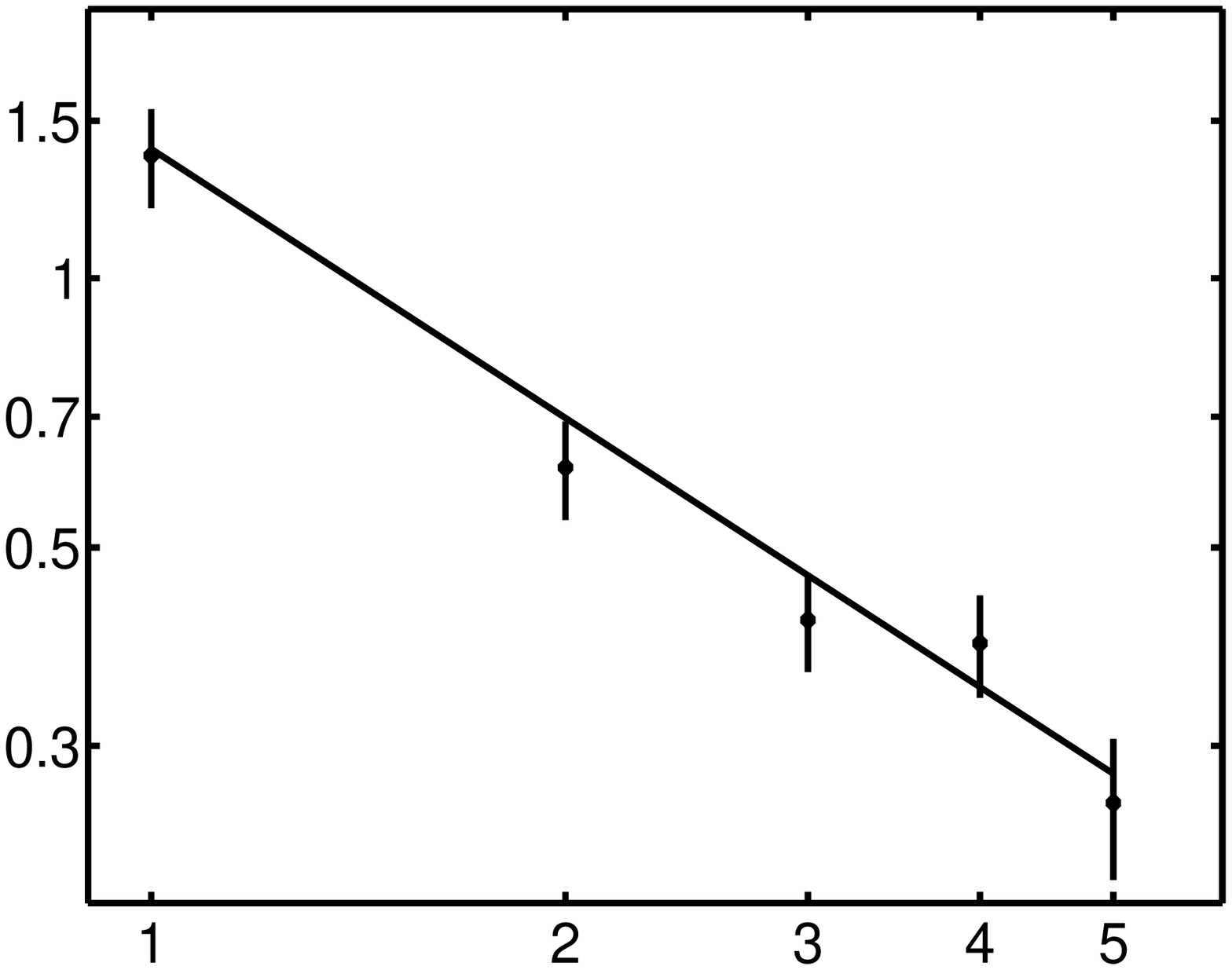}}
        \put(5.7,4.3){\mbox{$\phi(y)=y^2$, $\nu=2$, $\kappa_3=1.5$}}
        \put(4.2,2.5){\Large{$\frac{T^{\mbox{\tiny even}}_{\mbox{\tiny escape}}}{N^2}$}}
          \put(7.85,-0.6) {\mbox{$M$}}
         \end{picture} \end{center}
        \caption{Dependence of the escape time (time to learn at least one source signal) on the number of non-Gaussian sources $M$
         for Hebbian ICA with one projection ($K=1$). The solid line
    shows the slope predicted by the theory for comparison.
        The points with error bars denote the simulation results
        for $N=50$ averaged over 50 experiments.}
        \label{fig_esctime_m}
\end{figure*}

The shape of the potential for an example with two sources (with
equal kurtosis) and one projection ($K=1$, $M=2$) is shown on the
left of Figure~\ref{fig_potential}. In this case we have four
minimal potential barriers which the system has to overcome. For
the special case when all non-Gaussian sources have the same
kurtosis $\kappa^i_4=\kappa_4$ ($i=1,2, ...,M$), the escape time,
i.e. the mean first passage time for a single source to be
learned, is given by,
\begin{equation} \label{te_odd}
      T_{\mbox{\tiny escape}}^{\mbox{\tiny odd}}  \propto
    \frac{N^{3}}{2 M K}\exp\left[\frac{3\,\langle\phi^2(\mu)\rangle \, \nu}
    {8
    \,|\kappa_4\langle\phi'''(\mu)\rangle
    |}\right],
\end{equation}
 where $\sigma=\mbox{sign}(\langle \phi'''(\mu) \rangle \kappa_4)$ is a
 necessary condition for successful learning. Notice that the time-scale for escape diverges with
the learning rate. On the left of Figure~\ref{fig_esctime_m} we
show
 numerical simulations for this situation. We have generated data of
 dimension $N=50$ with $M$ non-Gaussian sources each with a uniform
 distribution and we extract a single projection $K=1$. The figure
 shows the dependence of the escape time on the number of non-Gaussian
 sources, on a log-log plot. The solid line shows the inverse
 scaling with $M$ predicted by the theory and this is consistent with the
 simulation results. A source was considered learned when
 the associated overlap $R_{ij}$ was observed to be greater than a
 threshold magnitude (0.1 greater than the position of the potential barrier).

If the sources have different kurtosis then the algorithm will be most
likely to learn the source with highest kurtosis first since this will
correspond to the escape point with the lowest potential barrier. The
mean first passage time will be dominated by the contribution from
this barrier for large learning rates and the escape time
will not depend on the number of sources.

\subsection{Even Non-linearity}

If the non-Gaussian signals are asymmetrical, then we can use
an even non-linearity, for example  $\phi(y)=y^2$. In this case
the appropriate scaling for the learning rate is $\eta=\nu
N^{-3/2}$.
 After
expanding Equation (\ref{eqn_dR}) near $\bm r=\bm 0$ we find that
the mean and covariance of the change in $\bm r$ at each iteration
are given by (to leading order in $N^{-1}$),
\begin{eqnarray}\label{mean_even}
   &&\mbox{E}[\Delta r_{ij}] =  ( -
    \mbox{$\frac{1}{2}$}  \langle\phi^2(\mu)\rangle \nu^2 r_{ij} +
    \mbox{$\frac{1}{2}$} \kappa^j_3\langle \phi''(\mu) \rangle \sigma_{ii} \nu
    r_{ij}^2
   -\mbox{$\frac{1}{2}$}\langle\phi(\mu)\rangle^2 \nu^2\sum^{K}_{l \neq i} r_{lj})N^{-2},\\
    &&\mbox{Cov}[\Delta
    r_{ij}, \ \Delta r_{kl}] =
    \langle \phi^2(\mu)\rangle \nu^2 \ \delta_{ik} \delta_{jl} N^{-2} ,
\label{cov_even}
\end{eqnarray}
where $\kappa^j_3$ is the third cumulant of the $j$-th source
distribution (third central moment), which measures skewness, and
brackets denote averages over Gaussian variables $\bm \mu
\sim{\cal N}(\bm 0, \bm I)$. Again the system can be described by
a Fokker-Planck equation for large $N$ but now with shorter
characteristic time-scale $(\delta t)^{-1}=N^2$. The system
is locally equivalent to a diffusion process in the cubic
potential
\begin{eqnarray}
    U(\bm r)  =\sum^{K}_{i=1}\sum^{M}_{j=1}(\mbox{$\frac{1}{4}$}\langle \phi^2(\mu) \rangle \nu^2 r_{ij}^2
    - \mbox{$\frac{1}{6}$}\kappa^j_3 \langle \phi''(\mu)\rangle \sigma_{ii}\nu \,
    r_{ij}^3
    + \mbox{$\frac{1}{2}$} \langle\phi(\mu)\rangle^2 \nu^2 \sum^{K}_{l \neq i}  r_{lj} r_{ij}) \ ,
\end{eqnarray}
with a diagonal diffusion matrix of magnitude
$D=\langle\phi^2(\mu)\rangle \nu^2$ as before. In this case the sum in the
potential contains ``cross-terms'' which depend on more than one
element in $\bm r$. The dynamics is therefore not equivalent to
the one-dimensional case and features of the potential will depend on
the particular value of $K$ and $M$ considered, making analysis less
straightforward than for the odd non-linearity.

\begin{figure*}[t!]
    \setlength{\unitlength}{0.95cm} \begin{center}
    \begin{picture}(9,3) \epsfysize = 15cm
    \put(-3,-0.2){
\epsfbox[50 0 550 600]{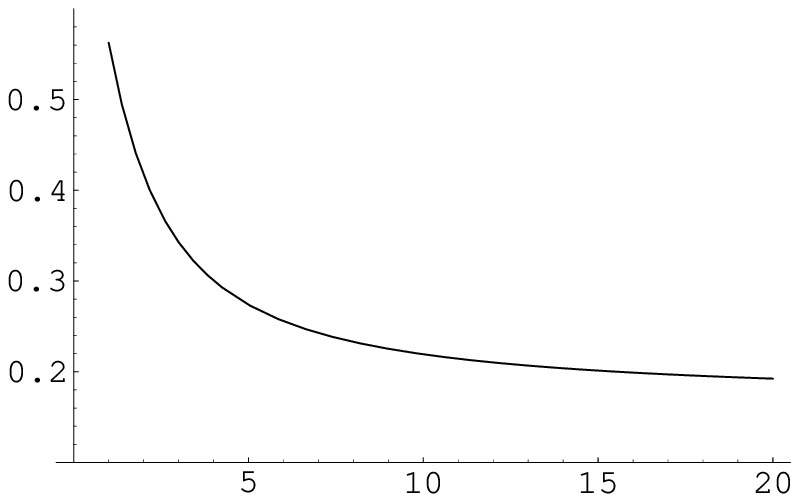}}
      \put(-2.1,3.5){\mbox{$f(K)$}}
    \put(1.3,-0.6){\mbox{$K$}}
        \epsfysize =6cm \put(5,-2.2){\epsfbox[50 0 550 600]{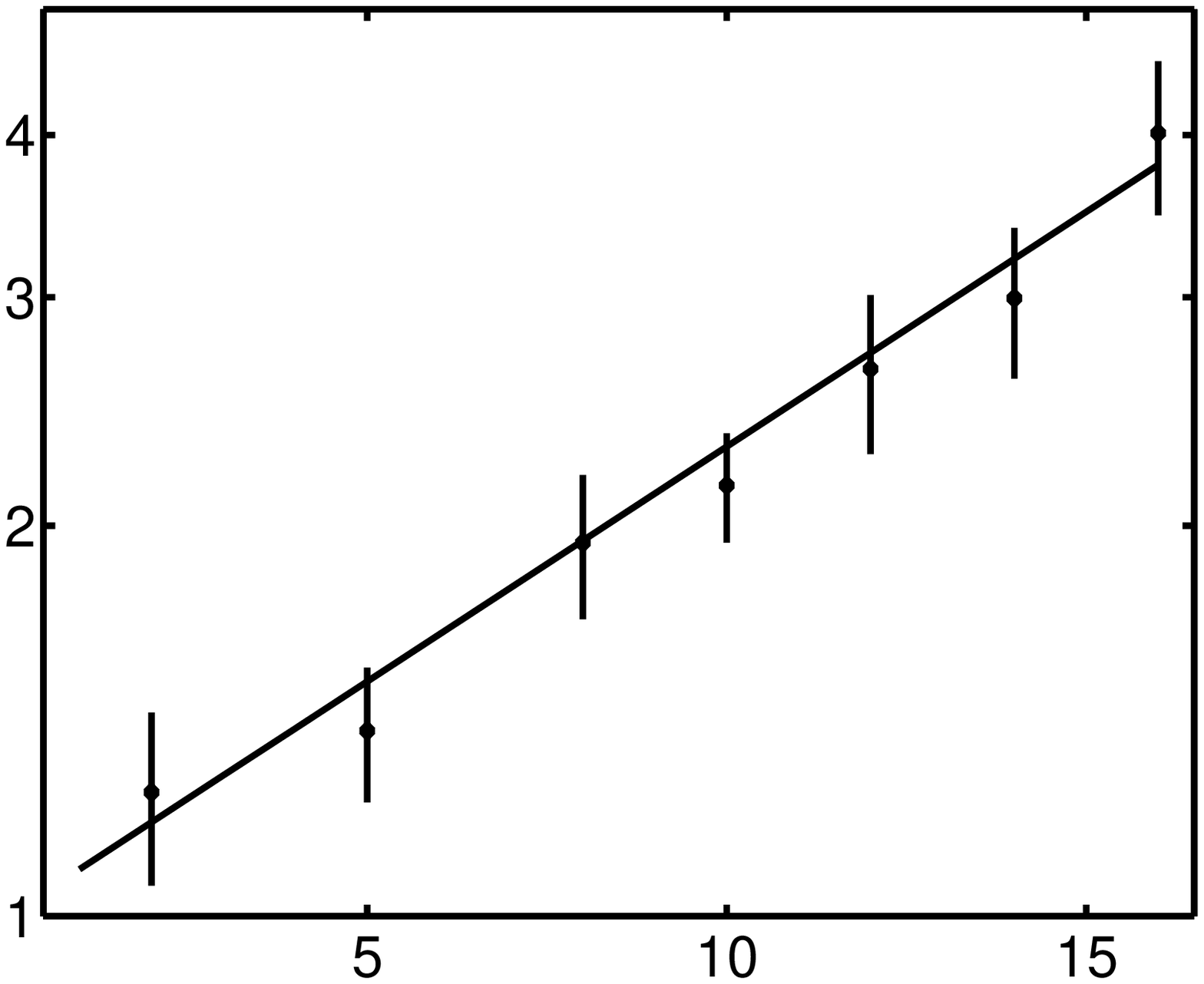}}
           \put(4.5 ,3){\Large{$\frac{T^{\mbox{\tiny even}}_{\mbox{\tiny escape}}}{N^2}$}}
        \put(8,-0.6){\mbox{$\nu^2$}}
         \end{picture} \end{center}
        \caption{Dependence of effective size of barrier of number of projections $K$
  for the specific choice of the even non-linear function $\phi(\mu)=\mu^2$ is shown on the left.
  Dependence of the escape time on learning rate $\nu$ for the one-dimensional Hebbian ICA
        ($K=M=1$) is shown on the right. The solid line shows the
    trend predicted by the theory.
        The points with error bars denote the simulation results
        for $\kappa_3=1.5$ and  $N=50$ averaged over 50 simulations.}  \label{fig_esctime_nu}
\end{figure*}

The shape of the potential for an example with two sources of equal
skewness and one projection ($K=1$, $M=2$)
is shown on the right of Figure~\ref{fig_potential}. In this case
we have a ledge in the potential with two points of minimum height
$\Delta U$ (escape points). In the general case we find that the
effective size of the minimal barriers is given by,
\begin{equation}\label{escapetime_even}
      \frac{\Delta U}{D}=\frac{f(K) \ \nu^2}{\langle\phi^2(\mu)\rangle  \
    (\kappa^i_3)^2}\ ,
\end{equation}
where the function $f(K)$ has a complex form which depends on the choice
of non-linear function $\phi(\mu)$. For example, the shape of this function for
$\phi(\mu)=\mu^2$ is shown on the left in
Figure~\ref{fig_esctime_nu}. We see that the size of potential
barrier decreases with increasing number of projections $K$ and this
appears to be a general feature of the function.
This suggests that parallel algorithms for extracting
asymmetrical signals may prove more efficient than deflationary
ones which separate one signal at a time.

For the case of a single projection ($K=1$) the potential does
decompose into a sum of independent terms and each component of
$\bm r$ evolves independently. For the case of sources with equal
skewness $\kappa^i_3=\kappa_3$ ($i=1,2, ...,M$), the mean first
passage time, i.e. the time until one of the source signals is
learned, is then given by,
\begin{equation}\label{te_1d}
T_{\mbox{\tiny escape}}^{\mbox{\tiny even}}  \propto
    \frac{N^{2}}{M}\exp\left[\frac{1}{12}\left(\frac{\langle\phi^2(\mu)\rangle\nu}{\kappa_3\langle\phi''(\mu)\rangle}\right)^2\right].
\end{equation}
Numerical results from simulations of this scenario with $N=50$
are shown on the right of Figures~\ref{fig_esctime_m} and
\ref{fig_esctime_nu}. The asymmetrical sources used in the
simulations are binary and each have the same skewness with
$\kappa_3 = 1.5$. In Figure~\ref{fig_esctime_m} we show the escape
time as a function of the number of sources $M$ on a log-log plot.
The solid line shows the inverse scaling predicted by the theory
and is consistent with the experimental results. On the right of
Figure~\ref{fig_esctime_nu} we show how the escape time (on a log
scale) depends on the learning rate parameter $\nu$. The slope of
the solid line is the theoretical prediction from
Equation~(\ref{te_1d}) and is consistent with the simulation
results.

\section{Transient Dynamics}
\label{sec_transient}

\begin{figure*}[t!]
    \setlength{\unitlength}{0.9cm} \begin{center}
    \begin{picture}(9,4) \epsfysize = 6cm
    \put(-2,-2.5){\epsfbox[50 0 550 600]{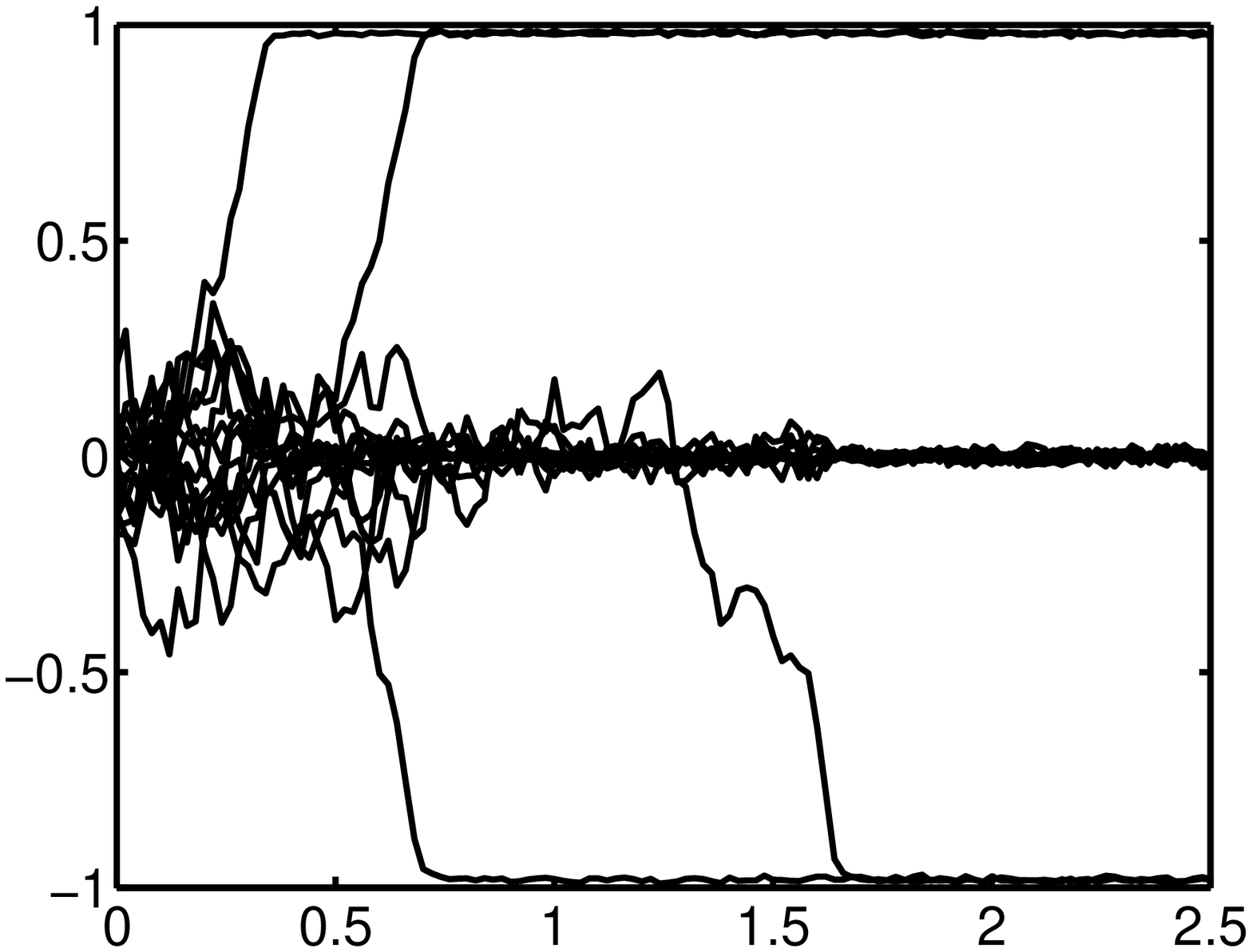}}
    \put(1.8,4.5){\mbox{$\phi(y)=y^3$,\ $\kappa_4=-4/9$, \ $\nu=1$}}
    \put(-2.5,2.5){\mbox{$R_{ij}$}}
    \put(-0.8,3.4){\mbox{\small{$R_{11}$}}}
    \put(-0.2,2.6){\mbox{\small{$R_{22}$}}}
    \put(-0,+0.4){\mbox{\small{$R_{33}$}}}
    \put(1.9,+0.4){\mbox{\small{$R_{44}$}}}
    \put(0.9,-0.6){\mbox{$t/N^3$}}
              \epsfysize =6cm \put(5,-2.5){\epsfbox[50 0 550 600]{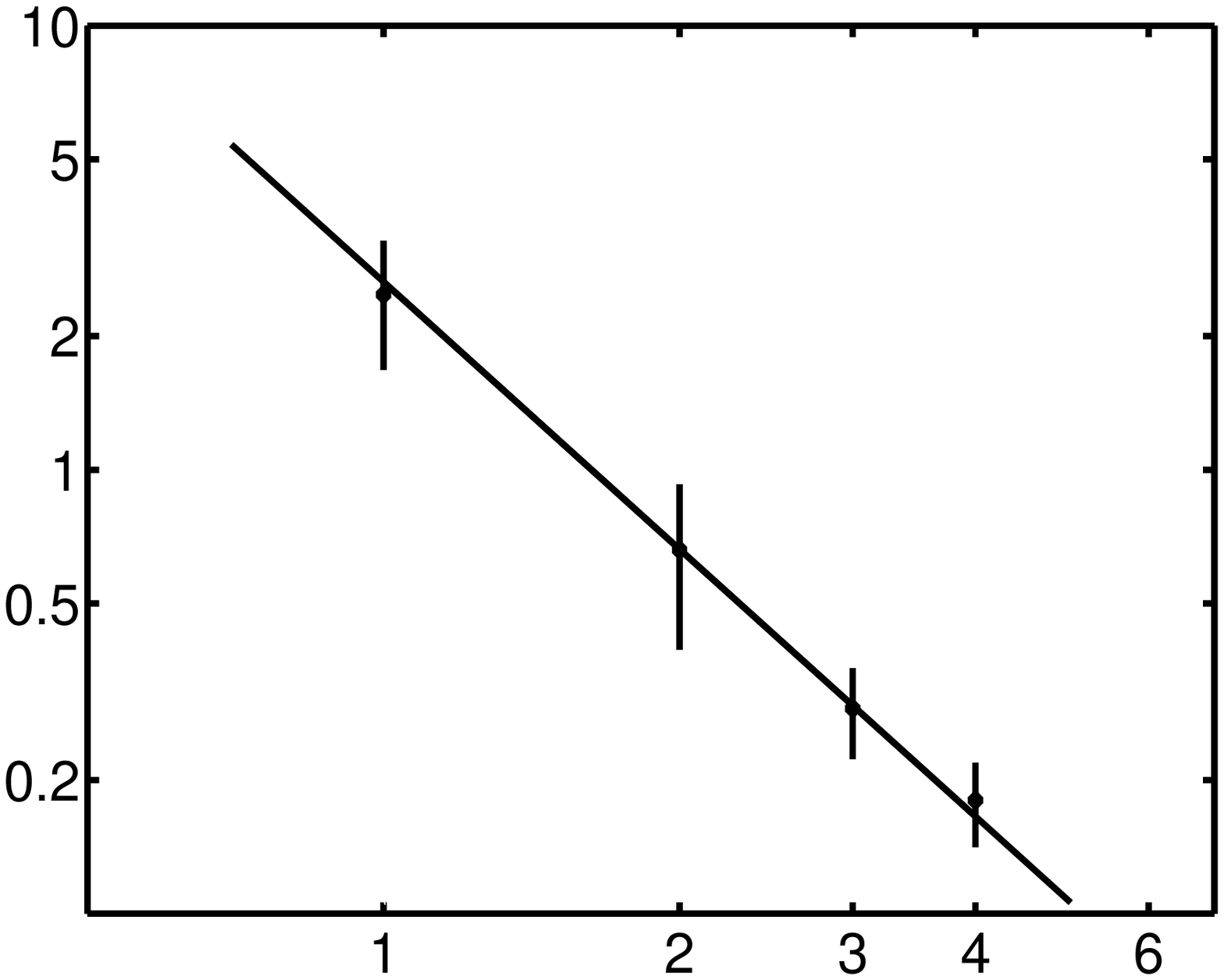}}
        \put(4.2,2.5){\Large{$\frac{\triangle t_i^{\mbox{\tiny escape}}}{N^3}$}}
          \put(5.8,-0.6) {\mbox{\small number of unlearned sources}}
         \end{picture} \end{center}
        \caption{\label{fig_5} Transient dynamics of the Hebbian ICA algorithm for $K=M=4$ and
        $N=50$.
        The left plot shows the simulation results for a single
        run. Dependence of time required to learn next source signal on the number of
        unlearned sources is shown on the right plot using a log-log scale. The solid line
    shows the trend predicted by the theory.
        The points with error bars denote the simulation results
        averaged over 10 simulations.
    }
\end{figure*}

Consider the more general situation when $T<M$ non-Gaussian
sources have already been learned by the system. The corresponding
fixed points of Equation~(\ref{eqn_dR}) are $R^*_{ij}
=\delta_{ij}\ \mbox{I}\left[i \leq T\right]$, where
$\mbox{I}\left[i \leq T\right]$ is defined by (\ref{predicate}).
We have already considered the special case when $T=0$ which is
appropriate close to the initial conditions when no sources have
yet been learned. A typical learning dynamics proceeds by passing
through these states one by one until all the sources are learned.
We show such a dynamical trajectory on the left of
Figure~\ref{fig_5}. The indices have been labeled retrospectively
so that the sources are learned in order although this labeling is
clearly arbitrary and only chosen for notational convenience. It
appears that the typical time to learn each source increases as
more sources are learned, a phenomenon which is explained below.

We introduce new $O(1)$ scaled variables,
\begin{equation}\label{vi}
  \nu=\eta \ N^{d}, \ \ \bm v =(\bm R-\bm R^*)\sqrt{N},
\end{equation}
where $N$ is the input dimension and $d$ is the scaling order for
the learning rate. For the case of an even non-linearity we set
$d=\frac{3}{2}$ while for the case of an odd non-linearity we
choose $d=2$. In our new variables the fixed point is $\bm v = \bm
0$. We can compute the mean and covariance of these variables as
we did for the $\bm r$ variables close to the initial conditions
in the previous Section. In the present case it is convenient to
consider four categories of variables separately.

\begin{enumerate}

\item{$i\leq T, \ j \leq T$.}
\begin{eqnarray}
\mbox{E}[\Delta v_{ij}] & = &\left[-\left(\xi_i +
    \mbox{$\frac{1}{2}$}\xi_j\right) v_{ij} -\mbox{$\frac{1}{2}$}\
    \xi_i v_{ji} \right] \nu N^{1-p} + \ O(N^{-p}) \ , \nonumber \\  \mbox{Cov}[\Delta
    v_{ij}, \ \Delta v_{kl}] & = & O(N^{-p}) \ .
 \label{dvij1}
\end{eqnarray}

\item{$i > T , \ j \leq T$}
\begin{equation}
\mbox{E}[\Delta v_{ij}] =  -\mbox{$\frac{1}{2}$}\xi_j v_{ij} \nu N^{1-p}
    + \ O(N^{-p}) \ , \quad \mbox{Cov}[\Delta
    v_{ij}, \ \Delta v_{kl}] = O(N^{-p}) \ .
 \label{dvij2}
\end{equation}

\item{$i \leq T , \ j > T$}
\begin{equation}
\mbox{E}[\Delta v_{ij}] =  -\xi_i v_{ij} \nu N^{1-p}
    + \ O(N^{-p}) \ , \quad \mbox{Cov}[\Delta
    v_{ij}, \ \Delta v_{kl}] = O(N^{-p}) \ .
 \label{dvij3}
\end{equation}

\item{$i > T , \ j > T$}
\begin{subequations}
\begin{gather}
\mbox{E}[\Delta v_{ij}]= -\frac{\partial U(\bm
  v)}{\partial  v_{ij}}N^{ -p} + O(N^{-p-1})\ ,\\
  \mbox{Cov}[\Delta
    v_{ij}, \ \Delta v_{kl}] = D \ \delta_{ik}\delta_{jl} N^{-p} + O(N^{-p-1})\ .
 \label{dvij4}
   \end{gather}
\end{subequations}

\end{enumerate}
In these equations, $p$ is the scaling order of our system with $p=2$
for the even non-linearity and $p=3$ for the odd non-linearity. We
define,
\[
    \xi_i = \sigma_{ii}\left(\mbox{E}_{s_i}[s_i \phi(s_i) -
    \phi'(s_i)]\right) \ .
\]
In Equation~(\ref{dvij4}) the exact expressions for the potential
$U(\bm v)$ and for the diffusion coefficient $D$ have the same
form as those which were found in Section \ref{sectionIV} given by
Equations~(\ref{mean_odd}), (\ref{cov_odd}), (\ref{mean_even}) and
(\ref{cov_even}).

In the first three groups of variable we observe that the
fluctuations are of negligible order, so that the dynamics can be
described by linear differential equations in the large $N$ limit
with a relatively fast time-scale of $\delta t^{-1}=N^{p-1}$.
Equations~(\ref{dvij1}), (\ref{dvij2}) and (\ref{dvij3}) therefore
converge exponentially to the fixed point as long as the condition
in Equation~(\ref{stab}) is met. However, the variables in the
fourth group display a similar diffusive dynamics to that
considered in the previous Section. The dynamics for these
variables is completely equivalent to the $\bm r$-dynamics close
to the initial conditions. Therefore we observe the same behaviour
in these variables, with localisation at the fixed point until one
component escapes over the potential barrier resulting in another
source being learned. Once all the sources are learned we
effectively have $T=M$ and only the first three groups of
variables above remain. In this case we can increase the learning
rate to an $O(N^{-1})$ quantity in principle without the
stochastic effects dominating, but then the learning rate should
be annealed as described by \citet{rattray02} in order to converge
asymptotically to the optimal solution.

The picture on the left in Figure \ref{fig_5} is a good
illustration of the transient dynamics. We show numerical
simulations for the typical dynamics of a Hebbian ICA algorithm
extracting four ($K=M=4$) uniformly distributed non-Gaussian
sources from an $N=50$ dimensional data set. On the right we show a log-log plot of the time $\triangle
t_i^{\mbox{\tiny escape}}$ required to learn the next source
signal in the case when $i$ non-Gaussian sources have already
been learned by the system and we see that it is consistent with the
expected trend shown by the solid line. The total learning time for
extracting all the non-Gaussian sources in this case will be
\begin{equation}\label{te_total}
  T_{\mbox{\tiny escape}}^{\mbox{\tiny \ total}}=\sum_{i=0}^{M-1} \triangle t_i^{\mbox{\tiny escape}}
  \propto \exp\left[\frac{\Delta U}{D}\right]\sum_{i=0}^{M-1} \frac{1}{2 (M-i)^2}\ ,
\end{equation}
where $\Delta U/D$ is given by Equation~(\ref{Afactor}).

\section{Natural Gradient ICA}
\label{sec_nat}

Natural gradient algorithms have been developed for ICA which use
the structure of the parameter space to define a Riemannian
gradient descent direction~\citep{amari96}. Along with closely
related relative gradient algorithms \citep{cardoso96} these
methods provide some
advantages over standard gradient descent methods, such as greater
simplicity, robustness and asymptotic efficiency~\citep{amari98}.
However, these algorithms have mainly been defined for the special
case where the mixing matrix is square and invertible.

The algorithm we use here searches the space of tall thin
orthogonal matrices. This allows it to extract a relatively small
number of independent components from a high-dimensional data set
possibly containing Gaussian components. Standard natural gradient
ICA algorithms are not appropriate in this case and we therefore
need a different approach. One possibility would be to use a
parameterisation of the set of orthogonal matrices. This approach
is considered by \citet{moon02} who provide an interesting
reinterpretation of natural gradient as a pullback. This allows
them to define natural gradient algorithms for various structured
matrices. Although they restrict their attention to square,
invertible matrices their ideas could be extended to tall thin
matrices. However, the available parameterisations appear to be
quite complex in this case and computing the gradient even more
so.

\begin{figure*}[t]
    \setlength{\unitlength}{0.9cm} \begin{center}
    \begin{picture}(9,4) \epsfysize = 6cm
    \put(-2,-2.5){\epsfbox[50 0 550 600]{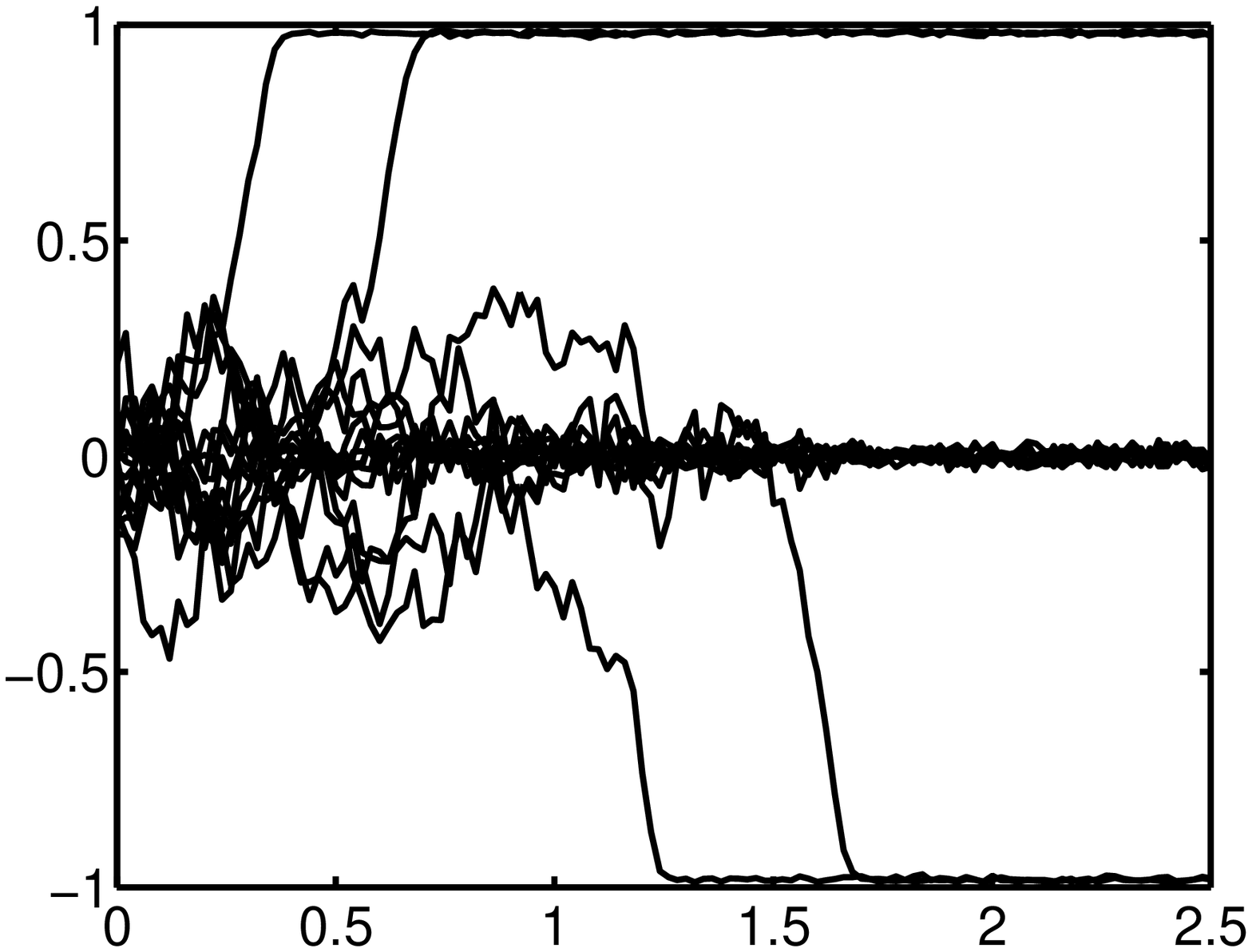}}
    \put(1.8,4.4){\mbox{$\phi(y)=y^3$,\ $\kappa_4=-4/9$, \ $\nu=1$}}
    \put(-2.5,2.5){\mbox{$R_{ij}$}}
    \put(-0.8,3.4){\mbox{\small{$R_{11}$}}}
    \put(-0.15,2.9){\mbox{\small{$R_{22}$}}}
    \put(0.29,+0.4){\mbox{\small{$R_{33}$}}}
    \put(1.9,+0.4){\mbox{\small{$R_{44}$}}}
    \put(0.9,-0.7){\mbox{$t/N^3$}}
            \epsfysize =6cm \put(5,-2.5){\epsfbox[50 0 550 600]{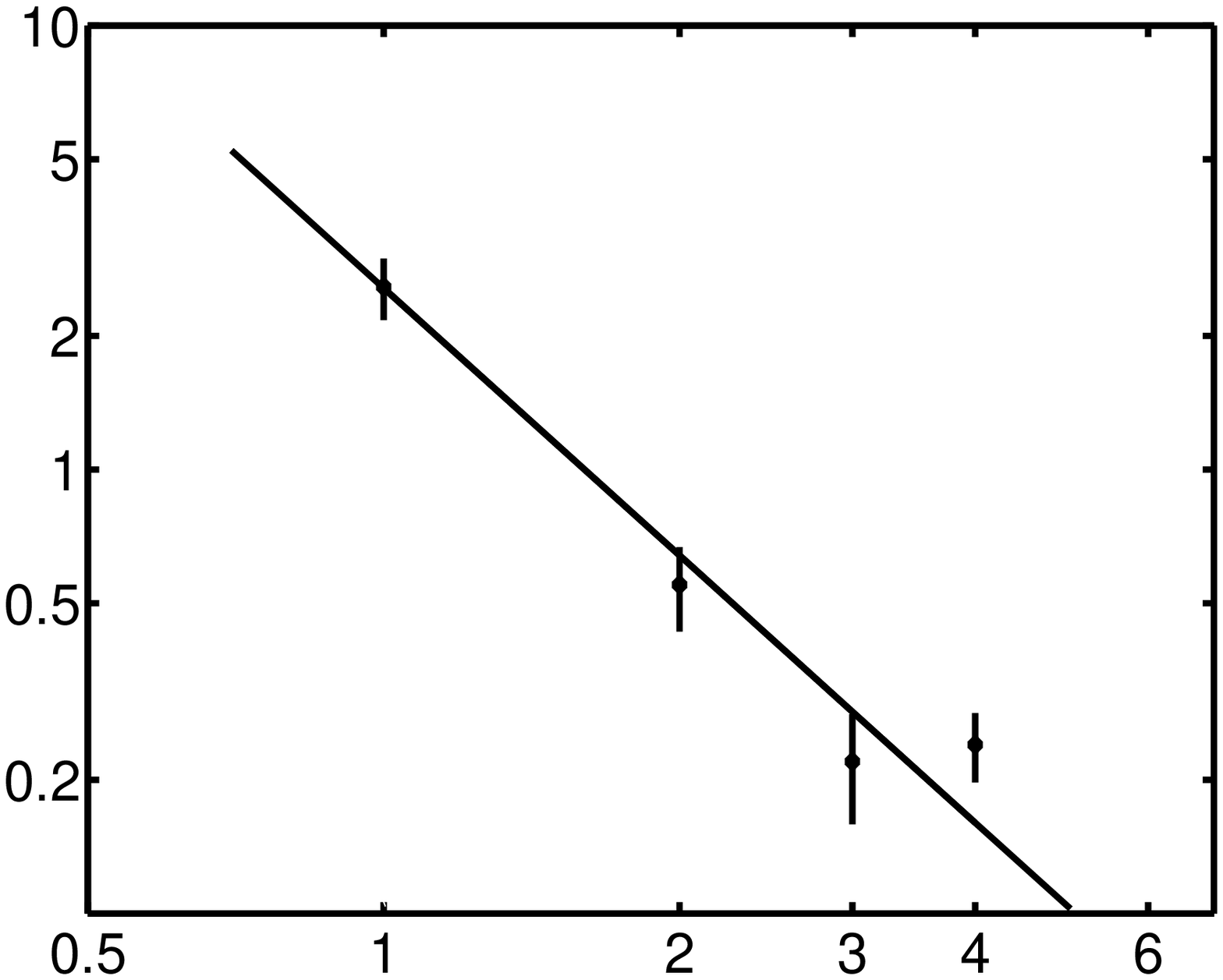}}
        \put(4.2,2.5){\Large{$\frac{\triangle t_i^{\mbox{\tiny escape}}}{N^3}$}}
          \put(5.8,-0.7) {\mbox{\small number of unlearned sources}}
         \end{picture} \end{center}
        \caption{\label{fig_ngd} Transient dynamics of the natural gradient Hebbian ICA algorithm for $K=M=4$ and
        $N=50$ (compare with Figure~\ref{fig_5}).
        The left plot shows the simulation results for a single
        run. Dependence of time required to learn next source signal on the number of
        unlearned sources is shown on the right plot using a log-log scale. The solid line
    shows the trend predicted by the theory.
        The points with error bars denote the simulation results
        averaged over 10 simulations.
    }
\end{figure*}

The approach of \citet{moon02} is to use a set of coordinates
which are intrinsic to the manifold. An alternative approach is to
use the original variables subject to constraints, i.e. work in
the space of tall thin matrices $\bm W$ but impose the
orthogonality constraint,
\begin{equation}\label{wortogonal}
 \bm W^\tr \bm W =\bm I \ .
\end{equation}
This is the approach taken by \citet{edelman99geometry} and it
leads to a much more straightforward gradient definition for ICA which
is described by~\citet{amari1999}. The
constraint surface is known as a Stiefel manifold and for a
function $F(\bm W)$ defined on the Stiefel manifold, the
``natural'' gradient of $F$ at  the point $\bm W$ of the manifold is
defined by
\begin{equation}\label{ngstiefel}
  \tilde{\bm \nabla}_W F = \bm \nabla_W F - \bm W \bm \nabla_W F ^\tr \bm
  W  ,
\end{equation}
where the standard gradient $\bm \nabla_W F$ is the $K$-by-$M$
matrix of partial derivatives of $F$ with respect to the elements
of $\bm W$. The loss function used in Hebbian ICA is some non-quadratic
function of the projections $F(\bm y)$ and the standard gradient of
this function is given by
\begin{equation}\label{normalgrad}
\bm \nabla_W F = \bm x \phi(\bm y)^\tr \bm \sigma \ .
\end{equation}
Then, according to Equation~(\ref{ngstiefel}), the natural
gradient of this function on the Stiefel manifold will
be~\citep{amari1999},
\begin{equation}\label{ng}
\tilde{\bm \nabla}_W F = \bm x \phi(\bm y)^\tr \bm \sigma - \bm W
\bm \sigma \phi(\bm y)\bm y^\tr .
\end{equation}
A disadvantage of using non-intrinsic variables is that the algorithm
is not guaranteed to stay on the manifold. This is especially
problematic for stochastic gradient algorithms which only
approximately follow the gradient direction. We therefore add the
same orthogonalisation term used in the standard Hebbian
algorithm. The natural gradient algorithm then has the following
form,
\begin{equation}\label{NGdW}
 \bm \Delta \bm W = \eta \, [\bm x
\bm\phi(\bm y)^\tr \bm \sigma - \bm W \bm \sigma \phi(\bm y)\bm
y^\tr ] + \alpha \bm W(\bm I - \bm W^\tr\bm W) \ .
\end{equation}
The update increment for the overlap matrices
$\bm R\equiv \bm W^\tr\bm A_s$ and $\bm Q \equiv \bm W^\tr\bm W$
at every learning iteration is found to be,
\begin{eqnarray}
    \Delta \bm R & = & \eta \left(\bm \sigma \bm\phi(\bm y)\bm s^\tr  - \bm y \phi(\bm y)^\tr\bm \sigma \bm R\right)+ \alpha (\bm I-\bm Q)\bm R \ , \label{ngdr} \\
    \Delta \bm Q & = & \eta \bm \sigma (\bm I - \bm Q + \alpha(\bm I - \bm
    Q)-\alpha(\bm I - \bm Q)\bm Q)\bm\phi(\bm y)\bm y^\tr + \alpha^2(\bm I - \bm
    Q)^2\bm Q \nonumber \\
  &  +& \eta\bm\sigma\bm y\bm\phi(\bm
    y)^\tr(\bm I -\bm Q + \alpha(\bm I - \bm Q)-\alpha \bm Q (\bm I - \bm Q))  + 2\alpha(\bm I - \bm Q)\bm Q \nonumber
 \\ & + & \eta^2\bm\phi(\bm y)\bm x^\tr\bm x\bm\phi(\bm y)^\tr
 \ .
     \label{ngdq}
\end{eqnarray}
After adiabatic elimination of the $\bm Q$ variables by a similar
procedure as we carried out for the case of Hebbian ICA (see
Section \ref{sectionIII}) we have the following dynamical
equations for the overlaps,
\begin{eqnarray}
 \Delta \bm R = \eta  \bm\sigma\left( \bm\phi(\bm
y)\bm s^\tr - \bm y\bm\phi(\bm y)^\tr\bm R\right)
 - \mbox{$\frac{1}{2}$} \eta^2 N \bm\phi(\bm
y)\bm\phi(\bm y)^\tr \bm R\ . \label{eqn_ngdR}
\end{eqnarray}
The typical learning dynamics of this natural gradient version of
Hebbian algorithm is shown on the left of Figure~\ref{fig_ngd}, where we used
the odd non-linearity $\phi(y)=y^3$ to extract
        four symmetrical sources with kurtosis
    $\kappa_4=-4/9$ from 50-dimensional data ($K=M=4$, $N=50$).

Following the procedure outlined in Section~\ref{sec_transient} we
can expand near the general fixed points with $T\leq M$ sources
learned. Using the same variables we expand around $\bm v=\bm 0$
and find the following results
\begin{enumerate}

\item{$i\leq T, \ j \leq T$.}
\begin{equation}
\mbox{E}[\Delta v_{ij}] = \left[-\left(\xi_i + \xi_j\right)
     \right] v_{ij}\nu N^{1-p} + \ O(N^{-p}) \ , \quad \mbox{Cov}[\Delta
    v_{ij}, \ \Delta v_{kl}] = O(N^{-p}) \ .
 \label{nat_dvij1}
\end{equation}

\item{$i > T , \ j \leq T$}
\begin{equation}
\mbox{E}[\Delta v_{ij}] =  -\xi_j v_{ij} \nu N^{1-p}
    + \ O(N^{-p}) \ , \qquad \mbox{Cov}[\Delta
    v_{ij}, \  \Delta v_{kl}] = O(N^{-p}) \ .
 \label{nat_dvij2}
\end{equation}

\item{$i \leq T , \ j > T$}
\begin{equation}
\mbox{E}[\Delta v_{ij}] =  -\xi_i v_{ij} \nu N^{1-p}
    + \ O(N^{-p}) \ , \quad \mbox{Cov}[\Delta
    v_{ij}, \ \Delta v_{kl}] = O(N^{-p}) \ .
 \label{nat_dvij3}
\end{equation}

\item{$i > T , \ j > T$}
\begin{subequations}
\begin{gather}
\mbox{E}[\Delta v_{ij}]= -\frac{\partial U(\bm
  v)}{\partial  v_{ij}}N^{ -p} + O(N^{-p-1})\, ,\\
  \mbox{Cov}[\Delta
    v_{ij}, \  \Delta v_{kl}] = D \ \delta_{ik}\delta_{jl} N^{-p} + O(N^{-p-1}) .
 \label{nat_dvij4}
   \end{gather}
\end{subequations}

\end{enumerate}
As before, $p=2$ for the even non-linearity and  $p=3$ for the odd
non-linearity and the potential in the last case is the same as
for standard Hebbian ICA. We see that the equations for the first
set of variables ($i\leq T, \ j \leq T$) in
Equation~(\ref{nat_dvij1}) are simplified in comparison to
Equation~(\ref{dvij1}) and no longer contain cross-terms. This
means, for example, that the algorithm will enjoy uniform
asymptotic convergence if all sources have identical statistics
and $M=K$. Generally speaking the eigenvalues determining the
convergence of the variables in the first three groups have lower
variance and the asymptotic convergence of the natural gradient
algorithm will be faster than that of the Hebbian algorithm.
However, Equations~(\ref{nat_dvij4}) and (\ref{dvij4}) are
identical and therefore the transient dynamics of the algorithms
will be very similar. These are the variables which provide the
rate limiting factor and learning time-scale during the transient.

The plot on the right of Figure~\ref{fig_ngd} shows the escape time from each of the
transient fixed points encountered during the dynamics. These simulation
results confirm that the transient dynamics is very similar to the standard
Hebbian algorithm results (see Figure~\ref{fig_5}) as predicted by our theory.

\section{Conclusion}
\label{sec_conclusion}

The dynamics of on-line ICA learning has been studied in the limit of
large data dimension. We have analysed a Hebbian learning algorithm
which is appropriate for extracting a prescribed number of
components from high dimensional data possibly containing Gaussian
components. We also studied a natural gradient variant of the
algorithm which uses the gradient defined on the Stiefel manifold of
orthogonal matrices.

We find that the learning time-scale of both algorithms is mainly
determined by the transient dynamics. Learning takes place by a
sequence of symmetry breaking steps in which a new source is
learned and these steps can be described as a diffusion and escape
process. The learning time-scale is found to be longer than
expected from the analysis of related algorithms such as on-line
back-propagation and Sanger's PCA
algorithm~\citep[e.g.][]{saad95a,biehl98}. To learn each symmetric
source typically requires of the order of $N^3$ learning
iterations while to learn an asymmetrical source using an even
non-linearity typically requires of the order of $N^2$ learning
iterations. Both algorithms exhibit equivalent transient dynamics
and we only find an advantage in using the natural gradient
variant asymptotically.

\subsection*{Acknowledgments.} This work was supported by an EPSRC
award (ref. GR/M48123).

\bibliography{ica_preprint}

\begin{thebibliography}{25}
\expandafter\ifx\csname natexlab\endcsname\relax\def\natexlab#1{#1}\fi
\expandafter\ifx\csname url\endcsname\relax
  \def\url#1{{\tt #1}}\fi

\bibitem[Amari(1998)]{amari98}
S.-I. Amari.
\newblock Natural gradient works efficiently in learning.
\newblock {\em Neural Computation}, 10:\penalty0 251--276, 1998.

\bibitem[Amari(1999)]{amari1999}
S.-I. Amari.
\newblock Natural gradient learning for over- and under-complete bases in
  {ICA}.
\newblock {\em Neural Computation}, 11\penalty0 (8):\penalty0 1875--1883, 1999.

\bibitem[Amari et~al.(1996)Amari, Cichocki, and Yang]{amari96}
S.-I. Amari, A.~Cichocki, and H.~H. Yang.
\newblock A new learning algorithm for blind source separation.
\newblock In D.~S. Touretzky, M.~C. Mozer, and M.~E. Hasselmo, editors, {\em
  Neural Information Processing Systems 8}, pages 757--763. MIT Press,
  Cambridge, 1996.

\bibitem[Biehl(1994)]{biehl94}
M.~Biehl.
\newblock An exactly solvable model of unsupervised learning.
\newblock {\em Europhysics Letters}, 25:\penalty0 391--396, 1994.

\bibitem[Biehl and Schl{\"o}sser(1998)]{biehl98}
M.~Biehl and E.~Schl{\"o}sser.
\newblock The dynamics of on-line principle component analysis.
\newblock {\em Journal of Physics A}, 31:\penalty0 L97--L103, 1998.

\bibitem[Biehl and Schwarze(1995)]{biehl95}
M.~Biehl and H.~Schwarze.
\newblock Learning by on-line gradient descent.
\newblock {\em Journal of Physics A}, 28:\penalty0 643--656, 1995.

\bibitem[Cardoso and Laheld(1996)]{cardoso96}
J.-F. Cardoso and B.~Laheld.
\newblock Equivariant adaptive source separation.
\newblock {\em IEEE Transactions on Signal Processing}, 44:\penalty0
  3017--3030, 1996.

\bibitem[Edelman et~al.(1999)Edelman, Arias, and Smith]{edelman99geometry}
A.~Edelman, T.~A. Arias, and S.~T. Smith.
\newblock The geometry of algorithms with orthogonality constraints.
\newblock {\em SIAM Journal on Matrix Analysis and Applications}, 20\penalty0
  (2):\penalty0 303--353, 1999.

\bibitem[Engel and {Van den Broeck}(2001)]{engel01}
A.~Engel and C.~{Van den Broeck}.
\newblock {\em Statistical Mechanics of Learning}.
\newblock Cambridge University Press, 2001.

\bibitem[Gardiner(1985)]{gardiner}
C.~W. Gardiner.
\newblock {\em Handbook of Stochastic Methods}.
\newblock Springer-Verlag, New York, 1985.

\bibitem[Hyv{\"a}rinen(1999)]{hyva99}
A.~Hyv{\"a}rinen.
\newblock Survey on independent component analysis.
\newblock {\em Neural Computing Surveys}, 2:\penalty0 94--128, 1999.

\bibitem[Hyv{\"a}rinen and Oja(1997)]{hyva97}
A.~Hyv{\"a}rinen and E.~Oja.
\newblock A fast fixed-point algorithm for independent component analysis.
\newblock {\em Neural Computation}, 9:\penalty0 1483--1492, 1997.

\bibitem[Hyv{\"a}rinen and Oja(1998)]{hyva98}
A.~Hyv{\"a}rinen and E.~Oja.
\newblock Independent component analysis by general non-linear {Hebbian}-like
  learning rules.
\newblock {\em Signal Processing}, 64:\penalty0 301--313, 1998.

\bibitem[Kushner(1987)]{kushner87}
H.~J. Kushner.
\newblock Asymptotic global behaviour for stochastic approximation and
  diffusions with slowly decreasing noise effects: Global minimization via
  monte carlo.
\newblock {\em SIAM Journal of Applied Mathematics}, 47:\penalty0 169--185,
  1987.

\bibitem[Kushner and Clark(1978)]{kushner78}
H.~J. Kushner and D.~S. Clark.
\newblock {\em Stochastic Approximation Methods for Constrained and
  Unconstrained Systems}.
\newblock Springer-Verlag, New York, 1978.

\bibitem[Moon and Gunther(2002)]{moon02}
T.~K. Moon and J.~H. Gunther.
\newblock Contravariant adaptation on structured matrix spaces.
\newblock {\em Signal Processing}, 82\penalty0 (10):\penalty0 1389--1410, 2002.

\bibitem[Press et~al.(1992)Press, Teukolsky, and Vetterling]{crecipes}
W.~H. Press, S.~A. Teukolsky, and W.~T. Vetterling.
\newblock {\em Numerical Recipes in C}.
\newblock Cambridge University Press, 2nd edition, 1992.

\bibitem[Rattray(2002)]{rattray02}
M.~Rattray.
\newblock Stochastic trapping in a solvable model of independent component
  analysis.
\newblock {\em Neural Computation}, 14:\penalty0 421--435, 2002.

\bibitem[Rattray and Basalyga(2002)]{nips02-LT19}
M.~Rattray and G.~Basalyga.
\newblock Scaling laws and local minima in {Hebbian ICA}.
\newblock In T.~G. Dietterich, S.~Becker, and Z.~Ghahramani, editors, {\em
  Advances in Neural Information Processing Systems 14}, pages 495--501. MIT
  Press, Cambridge, 2002.

\bibitem[Rattray and Saad(1999)]{rattray99}
M.~Rattray and D.~Saad.
\newblock Analysis of natural gradient descent for multilayer neural networks.
\newblock {\em Physical Review E}, 59:\penalty0 4523--4532, 1999.

\bibitem[Rattray et~al.(1998)Rattray, Saad, and Amari]{rattray98}
M.~Rattray, D.~Saad, and S.-I. Amari.
\newblock Natural gradient descent for on-line learning.
\newblock {\em Physical Review Letters}, 81:\penalty0 5461--5464, 1998.

\bibitem[Saad(1998)]{saad98}
D.~Saad, editor.
\newblock {\em On-line learning in neural networks}.
\newblock Cambridge University Press, 1998.

\bibitem[Saad and Solla(1995)]{saad95a}
D.~Saad and S.~A. Solla.
\newblock Exact solution for on-line learning in multilayer neural networks.
\newblock {\em Physical Review Letters}, 74:\penalty0 4337--4340, 1995.

\bibitem[{van Kampen}(1992)]{kampen}
N.~G. {van Kampen}.
\newblock {\em Stochastic Processes in Physics and Chemistry}.
\newblock Elsevier, Amsterdam, 1992.

\bibitem[White(1989)]{white89}
H.~White.
\newblock Learning in artificial neural networks: A statistical perspective.
\newblock {\em Neural Computation}, 1\penalty0 (4):\penalty0 425--464, 1989.

\end{thebibliography}

\end{document}